\documentclass[11pt]{article}
\usepackage{graphicx}
\usepackage{epsfig}
\usepackage{graphicx,colordvi}
\usepackage[dvips]{color}
\usepackage{amssymb}
\usepackage{amsmath,amsfonts}
\usepackage{url}
\usepackage{bm}
\usepackage{longtable}
\def\a{\alpha}
\def\b{\beta}

\def\iomn{i\omega_n}
\def\inun{i\nu_n}
\def\vq{{\bf q}}
\def\vk{{\bf k}}

\def\vR{{\bf R}}

\def\t{\mbox{tr}}

\def\cG0{{\cal G}_0}
\def\GRRp{G^{\vR\vR'}}
\def\WRRp{W^{\vR\vR'}}
\def\GRR{G^{\vR\vR}}
\def\WRR{W^{\vR\vR}}
\def\cU{{\cal U}}
\def\cG{{\cal G}}
\def\GH{{G_H}}
\def \t2g{t$_{2g}$}

\begin{document}
\parskip 1ex

\begin{center}
\section{SCIENTIFIC HIGHLIGHT OF THE MONTH}
\end{center}

\vspace{0.3cm}
\rule{16.0cm}{1mm}
\vspace{2mm}
\setcounter{section}{0}
\setcounter{figure}{0}

\begin{center}

Dynamical Screening Effects in Correlated Electron Materials --
A Progress Report on Combined Many-Body Perturbation and Dynamical Mean 
Field Theory: ``GW+DMFT''

Silke Biermann

Centre de Physique Th\'eorique, CNRS UMR7644, Ecole Polytechnique, 91128
Palaiseau, France

\end{center}

\begin{abstract}
We give a summary of recent progress in the field of
electronic structure calculations for materials with
strong electronic Coulomb correlations. The discussion
focuses on developments beyond the by now well established 
combination of density functional and dynamical mean field 
theory dubbed ``LDA+DMFT''. It is organized around 
the description of {\it dynamical screening} effects
in the solid. Indeed, screening in the solid gives rise to
{\it dynamical} local Coulomb interactions $\mathcal{U}(\omega)$
\cite{PhysRevB.70.195104, PhysRevB.86.165105},
and this frequency-dependence leads to effects that cannot
be neglected in a truly first principles description
\cite{casula_effmodel}.
We review the recently introduced extension of LDA+DMFT
to dynamical local Coulomb interactions ``LDA+$\mathcal{U}(\omega)$+DMFT''
\cite{PhysRevB.85.035115, werner_Ba122}.
A reliable description of dynamical screening effects is
also a central ingredient of the ``GW+DMFT'' scheme 
\cite{gwdmft, ayral, ayral-prb}, a
combination of many-body perturbation theory in Hedin's
GW approximation and dynamical mean field theory.
Recently, the first GW+DMFT calculations including dynamical 
screening effects for real materials have been achieved, with 
applications to SrVO$_3$ \cite{jmt_svo, jmt_svo2} 
and adatom systems on surfaces 
\cite{hansmann}. We 
review these and comment on further perspectives in the field.
\\
This review is an attempt to put elements of the original works
\cite{PhysRevB.70.195104, PhysRevB.86.165105, 
casula_effmodel, PhysRevB.85.035115, werner_Ba122, 
gwdmft, ayral, ayral-prb, jmt_svo, hansmann}
into the broad perspective of the development of truly
first principles techniques for correlated electron materials.
\\
\
\\
NB. This review was written in June 2013. It does not yet include
the most recent development of ``Screened Exchange Dynamical Mean
Field Theory'' (A. van Roekeghem et al., Phys. Rev. Lett. {\bf 113} 266403
(2014)).
\\
\
\\
{\bf Cite as:} S. Biermann, J. Phys.: Condens. Matter {\bf 26} 173202 (2014).
\end{abstract}

\section{Electronic Correlations from First Principles?}

The electronic properties of solids are dominated by the
electronic states in the immediate proximity to the Fermi level.
This observation -- together with the insights from
renormalisation group techniques -- are the motivation
for the quest of low-energy effective models that describe
the physical phenomena taking place in condensed matter.
In the early days of correlated electron physics
models were most often phenomenologically motivated,
without the ambition of a microscopic derivation, let alone
a {\it quantitative} description. Within the last decade, however,
a new research field has developed at the interface of 
many-body theory and first prinicples electronic structure 
calculations. The aim is the construction of materials-specific
parameter-free many-body theories that preserve the {\it ab initio}
nature of density functional based electronic structure 
calculations, but incorporate at the same time a 
many-body description of Coulomb interactions beyond the
independent-electron picture into the description of
spectroscopic or finite-temperature properties.

Such ``correlation'' effects, that is, effects beyond the effective
one-particle picture, are indeed most striking in spectroscopic
probes, where they take the form of quasi-particle renormalisations 
or broadening due to finite lifetimes, and give rise to satellite 
features or atomic multiplets. An intrinsic temperature
dependence of the electronic structure of a metal, with a 
coherence-incoherence crossover delimiting Fermi liquid properties,
or a strongly temperature-dependent gap -- beyond what can be
explained by a Fermi factor -- are further hallmarks of electronic
correlations \cite{jmt_fesi}.

Historically, the first non-perturbative electronic structure 
techniques for correlated materials evolved from many-body 
treatments of the multi-orbital Hubbard Hamiltonian with
realistic parameters.
The general strategy of these so-called ``LDA++'' approaches 
\cite{Lichtenstein98, Anisimov97}
consists in the extraction of the 
parameters of a many-body Hamiltonian from first principles 
calculations and then solving the problem by many-body 
techniques. The procedure becomes conceptually involved,
however, through the need of incorporating effects of
higher energy degrees of freedom on the low energy part,
the so-called ``downfolding''.

For the one-particle part of the Hamiltonian, downfolding
techniques have been the subject of a vast literature
\cite{lowdin51,andersen00}, and are by now well established.
The task here is to define orbitals spanning the low-energy
Hilbert space of the electronic degrees of freedom of a 
solid in such a way that a low-energy one-particle Hamiltonian 
can be constructed whose spectrum coincides with the
low-energy part of the spectrum of the original one-particle 
Hamiltonian.\footnote{%
We do not enter here into details concerning the different
strategies of achieving such a construction: various frameworks,
such as muffin-tin orbitals methods \cite{andersen00}, 
maximally localised Wannier functions \cite{PhysRevB.56.12847}, 
or projected atomic orbitals \cite{aichhorn09} have been employed.}
Downfolding of the interacting part of a many-body
Hamiltonian is a less straightforward problem, which
has attracted a lot of attention recently.
The challenge is an accurate description of screening
of low-energy interactions by high-energy degrees of 
freedom. Indeed, the net result of the rearrangement
of the high-energy degrees of freedom as response to
a perturbation of the system is an effective reduction 
of the perturbation strength in the low-energy space.
It is for this reason that the effective Coulomb interaction
in a low-energy effective model for a correlated system
is in general an order of magnitude smaller than the
matrix element of the bare Coulomb interaction.
Nevertheless, the latter is recovered in the 
limit of high-frequencies of the perturbation, when
screening becomes inefficient.
The crossover -- as a function of frequency -- from the 
low-energy screened regime to the high-frequency
bare matrix element of $\frac{e^2}{|{\bf r} - {\bf r}^{\prime}|}$
takes place at a characteristic screening (plasma) frequency 
where the dielectric function exhibits a pole structure.

This frequency-dependence of the effective local
Coulomb interactions, the {\it dynamical Hubbard $\mathcal{U}(\omega)$}
and its consequences on the electronic structure of
correlated materials are at the center of the present review.
We first recall the formalism of the constrained random
phase approximation, as the simplest means to obtain
a quantitative estimate for the dynamical Hubbard interactions.
We then review a recent scheme to incorporate the dynamical
nature of the Hubbard interactions into dynamical mean field
based electronic structure calculations. This ``Bose factor
ansatz'' also gives a transparent physical interpretation 
of the observed new features, such as plasmon satellites
and renormalisations of spectral weight at low energies.
Electronic structure calculations with frequency-dependent
interactions -- using a scheme that should best be called
``LDA+$\mathcal{U}(\omega)$+DMFT'' -- 
for the iron pnictide compound BaFe$_2$As$_2$
illustrate the importance of these effects, while at the
same time revealing new unexpected many-body behavior
in the form of an incoherent (non-Fermi-liquid) regime.

Dealing with frequency-dependent interactions at the DMFT
level has been a major bottleneck in the
implementation of the combined ``GW+DMFT'' scheme since 
its proposal in 2003 \cite{gwdmft}. The recent advances concerning
this issue, both concerning Monte Carlo techniques and through
the Bose factor ansatz, have now unblocked the situation:
two calculations within GW+DMFT taking into account 
dynamical interactions have been achieved recently,
for SrVO$_3$ \cite{jmt_svo, jmt_svo2} and for systems of adatoms on surfaces
\cite{hansmann}.
We review these calculations, together with systematic
studies of an extended Hubbard model \cite{ayral}, which demonstrate
how the GW+DMFT scheme enables an additional type
of ``downfolding'': effects of long-range interactions
can in fact be ``backfolded'' into a purely local
effective quantity, a generalised Hubbard $\mathcal{U}(\omega)$,
which acquires its frequency-dependence due to screening
by non-local processes. The strength of these 
screening processes are shown to be strongly system-dependent
when the true long-range nature of Coulomb interactions is
taken into account, while simple rules of thumb work relatively
well in the case of an extended Hubbard model with nearest-neighbor
interactions only.

Finally, we define current open questions and comment on
further perspectives in the field.

\section{Calculating effective local Coulomb interactions from
first principles: from
``Hubbard U'' to dynamical Hubbard $\mathcal{U}(\omega)$}

We now turn to a mathematical formulation of 
the above described frequency-dependent behavior of the
effective Coulomb interactions 
(``Hubbard U'') to be used within low-energy
effective descriptions. The simplest way to describe the
frequency-dependence stemming from downfolding higher energy
degrees of freedom is given by the
``constrained random phase approximation'' (cRPA) technique
\cite{PhysRevB.70.195104}.
The cRPA provides an (approximate) answer to the following
question: given the Coulomb Hamiltonian in a large Hilbert
space, and a low-energy Hilbert space that is a subspace of
the former, what is the effective {\it bare} interaction to
be used in many-body calculations dealing only with the
low-energy subspace, in order for physical predictions for
the low-energy Hilbert space to be the same for the two
descriptions?
A general answer to this question not requiring much less
than a full solution of the initial many-body problem, the
cRPA builds on two approximations: it assumes (i) that the
requirement of the same physical predictions be fulfilled
as soon as in both cases the same estimate for the fully
screened Coulomb interaction, Hedin's $W$, is obtained
and (ii) the validity of the random phase approximation
to calculate this latter quantity.

The cRPA starts from a decomposition of the polarisation
of the solid in high- and low-energy parts, where the
latter is defined as given by all screening processes
that are confined to the low-energy subspace. The former
results from all remaining screening processes:
\begin{eqnarray} 
  P^{\rm high}=P - P^{\rm low},
\label{polsep}
\end{eqnarray}
One then calculates a partially screened interaction
\begin{eqnarray} \label{wrest}
  W^{\rm partial}(1,2) &\equiv& 
\int d3 \, \varepsilon_{\rm partial}^{-1}(1,3) v(3,2).
\end{eqnarray}
using the {\it partial} dielectric function
\begin{eqnarray} \label{dielectric}
  \varepsilon_{\rm partial}(1,2) &=& \delta(1-2) - 
\int d3 \, P^{\rm high}(1,3)v(3,2).
\end{eqnarray}
Here, the numbers represent space and time coordinates 
in a shorthand notation.

Screening $W^{\rm partial}$ by processes that live within the 
low-energy space recovers the fully screened interaction $W$.
This justifies the interpretation of the matrix elements of 
$W^{\rm partial}$ in a localized Wannier basis as the interaction 
matrices to be used as bare Hubbard interactions within a low-energy
effective Hubbard-like Hamiltonian written in that Wannier basis.

Hubbard interactions -- obtained as the static ($\omega=0$) limit 
of the onsite matrix element
$\langle |W^{\rm partial} | \rangle$ within cRPA -- have by now been
obtained for a variety of systems, ranging from transition
metals \cite{PhysRevB.70.195104} to oxides
\cite{miyake:085122,PhysRevB.74.125106,jmt_mno, PhysRevB.86.165105},
pnictides \cite{miyake2008, nakamura, miyake2010, imada2008}, or 
f-electron compounds \cite{jmt_cesf}, and several implementations
within different electronic structure codes and basis sets
have been done, e.g. within 
linearized muffin tin orbitals \cite{PhysRevB.70.195104}, maximally
localized Wannier functions \cite{miyake:085122, cRPA-friedrich, nakamura}, or
localised orbitals constructed from projected atomic
orbitals \cite{PhysRevB.86.165105}.
The implementation into the framework of the Wien2k package
\cite{PhysRevB.86.165105} made it possible that Hubbard $U$'s be calculated
for the same orbitals as the ones used in subsequent LDA+DMFT 
calculations, see e.g.~\cite{martins}.
Systematic calculations investigating the basis set dependence
for a series of correlated transition metal oxides revealed
furthermore interesting trends, depending on the choice of the
low-energy subspace. In contrast to common belief until then,
Hubbard interactions increase for example with the principal quantum 
number when low-energy effective models encompassing only the t$_{2g}$
orbitals are employed. These trends can be rationalised by two
counteracting mechanisms, the increasing extension of the orbitals
with increasing principal quantum number and the less efficient
screening by oxygen states \cite{PhysRevB.86.165105}.
\\

\begin{figure}[t!h]
\centering
\includegraphics[clip=true,
width=0.69\textwidth,angle=270]
{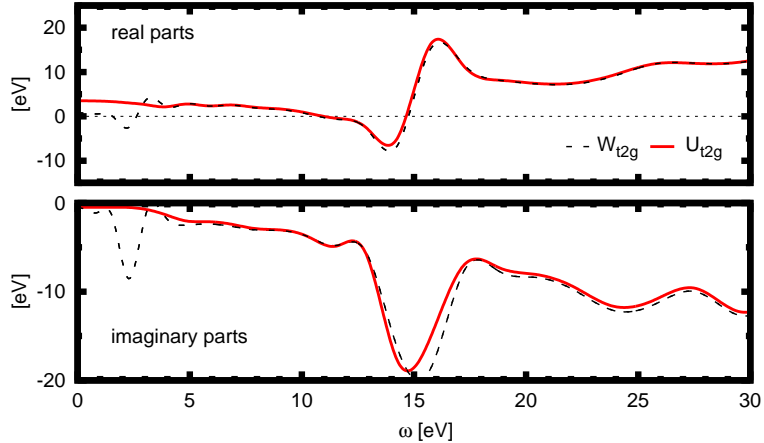}
\caption[bla]{Screened Coulomb interaction $W$
and partially screened Coulomb interaction $U$ for SrVO$_3$:
real (top panel) and imaginary (bottom panel) parts of the
matrix elements of the fully screened Coulomb interaction
$\langle R m R m| W(\omega) | R m R m \rangle$
and the partially screened one,
$\mathcal{U}=\langle R m R m| W_r(\omega) | R m R m \rangle$, 
within t$_{2g}$
Wannier functions $\phi_m(r-R)$, $m$=xy, yz, or xz,
centred on an atom at position $R$,
as calculated within the cRPA. 
Note the small low energy value of $\mathrm{Re} U(0)$, compared
to the matrix element of the bare Coulomb interaction
$\mathrm{Re} U(\infty)$, and the plasmon excitation at 15 eV. 
At high energies, screening by processes taking place
within the t$_{2g}$ manifold become irrelevant, and $W$
and $W_r$ coincide. At low energies, however, $W$ displays
additional frequency-dependence as compared to $W_r$,
originating from processes within the t$_{2g}$ manifold.
These processes are cut out from the polarisation $P_r$
used for calculating $W_r$, according to the cRPA
prescription.
Adapted from \cite{jmt_svo} (see also \cite{jmt_svo}).
}
\label{fig:U-svo}
\end{figure}

In general, values obtained within the cRPA have for a long time been
believed to be slightly ``too small'', since quite systematically
not only constrained LDA techniques result in larger values but
also many-body calculations that fix the interactions in order to 
obtain agreement with experiments usually employ slightly larger 
values than those obtained within cRPA.
This puzzle has been recently solved \cite{PhysRevB.85.035115, casula_effmodel}:
the key was found to lie in the frequency-dependence of the interactions
leading to additional renormalisations of the one-body Hamiltonian.
Indeed, as can be seen from Eq.~(\ref{wrest}), $W^{\rm partial}(\omega)$
is a function of frequency, and so are matrix elements derived
from it, in particular its local part, the Hubbard $\mathcal{U}(\omega)$. 
An example is given in Fig.~\ref{fig:U-svo}, where the diagonal
matrix elements $\langle R m R m| W(\omega) | R m R m \rangle$
and $U=\langle R m R m| W_r(\omega( | R m R m \rangle$ within t$_{2g}$
Wannier functions $\phi_m(r-R)$, $m$=xy, yz, xz,
centred on an atom at position $R$ are plotted
as a function of frequency.
The consequences of this dynamical nature of the effective
interactions are the subject of the following sections.

\section{Dynamical screening effects: plasmons, spectral
weight transfers and electronic polarons}

The explicit treatment of many-body problems with dynamical Hubbard 
interactions has by now become possible even in the realistic multi-orbital
case. This progress is due both to quite impressive advances in Monte Carlo
techniques and to the development of extremely accurate efficient 
approximations, 
that reduce the problem to a static one, at least in the antiadiabatic
limit when the characteristic screening frequencies are much larger
than other relevant energy scales of the problem (bandwidth and static 
Hubbard interaction $U(\omega=0)$). 
Several applications to materials have appeared, namely for SrVO$_3$
in \cite{PhysRevB.85.035115, jmt_svo, jmt_svo2, li-svo}, and to BaFe$_2$As$_2$ 
in \cite{werner_Ba122}.
Alternatively to a direct explicit treatment of the dynamical interactions,
in the antiadiabatic limit a mapping onto an effective low-energy
model with static interactions can also be performed, if only low-energy
properties, living on energy scales considerably smaller than the
plasma frequency, are of interest \cite{casula_effmodel}. 
This procedure is reviewed in Appendix A.

In the remainder of this section, we focus on the treatment of dynamical 
Hubbard interactions within the ``Bose factor ansatz'' (BFA),
which allows for a transparent physical interpretation of the
observed effects. Moreover, comparison of calculations for SrVO$_3$
within the BFA in \cite{PhysRevB.85.035115} and within Monte Carlo in
\cite{li-svo} demonstrate the impressive accuracy of this approach in the
regime relevant for real materials applications.

Extending the philosophy of the LDA+DMFT scheme to dynamically
screened interactions requires the use of a framework that allows
for a description of an explicit frequency-dependence of the interactions
$\mathcal{U}(\omega)$.
One possibility is to switch from the Hamiltonian
formulation of the ``LDA++'' approach to an action description where
the frequency-dependent nature of the interaction is readily
incorporated as a retardation in the interaction term  
\begin{eqnarray}
S_{int}[\mathcal{U}] 
= - \int_0^{\beta} \int_0^{\beta} 
d\tau d\tau^{\prime} \mathcal{U}(\tau- \tau^{\prime})
n(\tau) n(\tau^{\prime})
\end{eqnarray}
where we have assumed that the retarded interaction couples only
to the density $n(\tau)$.
Alternatively, it is possible to stick to a Hamiltonian formulation.
In order to describe the retardation effects in the interaction one
then needs to introduce additional bosonic degrees of freedom that
parametrise the frequency-dependence of the interaction.
Indeed, from a physical point of view, screening can be understood
as a coupling of the electrons to bosonic screening degrees of freedom
such as particle-hole excitations, plasmons or more
complicated composite excitations giving rise to shake-up satellites
or similar features in spectroscopic probes.
Mathematically, a local retarded interaction can be represented
by a set of bosonic modes of frequencies $\omega$ coupling to the 
electronic density with strength $\lambda_{\omega}$.
The total Hamiltonian 
\begin{eqnarray}
H= H_{LDA++} + H_{screening}
\end{eqnarray}
is then composed by a part of ``LDA++'' form
but with the local interactions given by the {\it unscreened}
local matrix elements of the bare Coulomb interactions $V$ and the
Hund's exchange coupling $J$ (assumed not to be screened by the
bosons and thus frequency-independent)
\begin{eqnarray}
H_{LDA++} &=& 
H^{KS}
+ \frac 12\sum_{imm^{\prime}\sigma } 
V_{mm^{\prime }}^in_{im\sigma}n_{im^{\prime }-\sigma }  
+ \frac 12\sum_{im\neq m^{\prime}\sigma} 
(V_{mm^{\prime }}^i-J_{mm^{\prime}}^i)n_{im\sigma }n_{im^{\prime }\sigma }  
\end{eqnarray}
and a screening part consisting of the local bosonic modes and their
coupling to the electronic density:
\begin{eqnarray}
H_{screening} = 
\sum_{i}\int d\omega \Big[\lambda_{i \omega}(b_{i\omega}^{\dagger} + b_{i\omega}) 
\sum_{m \sigma} n_{i m \sigma}+\omega b_{i\omega}^{\dagger}b_{i\omega}\Big].
\nonumber
\end{eqnarray}
As in standard LDA+DMFT, many-body interactions 
are included for a selected set of local orbitals, assumed
to be ``correlated''. The sums thus run over atomic sites $i$
and correlated orbitals $m$ centered on these sites.
Further, $H^{KS}$ represents a one-body Hamiltonian defined by
the DFT Kohn-Sham band structure, suitably corrected for double
counting terms. We show in Appendix B that despite of the bare
Coulomb interaction entering the Hamiltonian above, as a consequence
of the presence of the bosonic degrees of freedom, the standard
form of double counting with the {\it screened} interaction is
recovered.

Integrating out the bosonic degrees of freedom would lead back
to a purely fermionic action with retarded local interactions
\begin{eqnarray}
\mathcal{U}(\omega) = V + \int d\omega^{\prime} 
\lambda_{\omega^{\prime}}^2   
\left(\frac{1}{\omega- \omega^{\prime}}
-\frac{1}{\omega+ \omega^{\prime}} \right)
\end{eqnarray}
The above Hamiltonian thus yields a parametrisation of the
problem with frequency-dependent interactions provided that
the parameters are chosen as
$\text{Im} \mathcal{U}(\omega)=\pi\lambda_\omega^2$.
The zero-frequency (screened) limit is then given by
$U_0=V-2\int d\omega \frac{\lambda_{\omega}^2}{\omega} $.

The above form of the Hamiltonian corresponds to a multi-orbital
multi-mode version of the familiar Hubbard-Holstein Hamiltonian
describing a system of fermions coupled to bosonic modes.
The emergence of retarded interactions in the case of electron-phonon 
coupling has been investigated in detail for its
role in the BCS theory of pairing arising in conventional
superconductors.
In the current situation where the bosons represent plasmons
and other screening modes typical energy scales are radically
different, and the regime of importance is most often the
antiadiabtic one. Indeed, plasma frequencies of typical
transition metal based materials -- transition
metals themselves, their oxides, pnictides etc. -- are
of the order of 15 to 25 eV, whereas both the typical bandwidth
and static ($\omega=0$) Hubbard interaction are rather of the
order of a few eV. This hierarchy gives rise to a separation
of energy scales that enables a simple and transparent physical
interpretation of the solution of the Hubbard-Holstein Hamiltonian.

\begin{figure}[t]%
\centering
\includegraphics[scale=0.69]{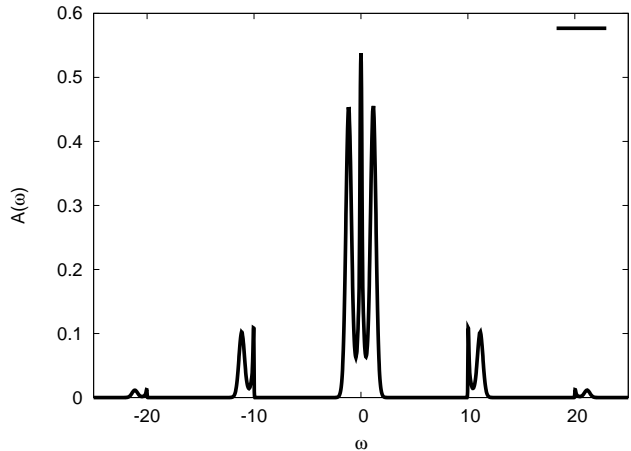}
\caption{Spectral function of the single-orbital single-mode
Hubbard Holstein model in the antiadiabatic limit. Replicae
of the low-energy part of the spectral function due to 
plasmon excitations are clearly seen. From \cite{PhysRevB.85.035115}.
}%
\label{cartoon}%
\end{figure}

We illustrate this fact on the most simple example, a half-filled 
{\it single-orbital} 
Hubbard-Holstein model 
with a {\it single local bosonic mode}
on a Bethe lattice with semi-circular
density of states. The frequency $\omega_0$ of the bosonic
mode is chosen to be the largest energy scale of the problem
so that the chosen parameter set places the system deep in the
antiadiabatic limit. The spectral function in this case is 
plotted in Fig. \ref{cartoon}.
It corresponds to a sequence of features located at energies
that are positive or negative multiples of the plasma 
frequency. They correspond to electron removal or addition 
processes where the (inverse) photoemission process itself is 
accompanied by the creation or annihilation of a certain 
number of screening bosons.
The low-energy part of the spectral function, close to the Fermi
level -- chosen to be the origin of energies -- is given by
electron removal or addition processes that do {\it not} 
change the number of screening bosons.
In the present simple half-filled case in a moderate correlation
regime, it displays a ``three-peak structure'', with
a central quasi-particle peak and upper and lower Hubbard bands,
typical of correlated metals.
Interestingly, however, even this part is modified by the
coupling to the bosons: indeed, since the full spectral function
is normalised, spectral weight appearing in plasmon replicae
of the main line reduces the weight contained in the latter.
The coupling to the bosonic degrees of freedom thus leads
to an additional mass renormalisation of the low-energy fermionic
degrees of freedom. This effect corresponds to the mass enhancement
due to the formation of ``electronic polarons'', fermions dressed
by their screening bosons just as usual polarons can be understood
as electrons dressed by the polarisation of the surrounding
lattice.
In the case of core level spectroscopies, such effects have
been extensively discussed, and the electron-boson couplings
above can be viewed as a local version of Hedin's ``fluctuation
potentials'' (albeit, in the cRPA sense, that is parametrising
not the fully screened interaction $W$ but rather the Hubbard
$U$) \cite{hedin}. 

When solving a many-body problem with static interactions -- to
simplify the notation we discuss here the case of a single-orbital
problem only  -- within
DMFT, the central ingredient is the solution of an effective local
problem (``impurity problem'') 
\begin{eqnarray}
S = 
- \int \int d\tau d\tau^{\prime} \sum_{\sigma} c_{\sigma}^{\dagger}(\tau^{\prime})
(\delta(\tau - \tau^{\prime}) \partial_{\tau} -
\Delta(\tau - \tau^{\prime})) c_{\sigma}(\tau)
+ \int d\tau U n_{\uparrow}n_{\downarrow}
\end{eqnarray}
with an effective local {\it bath} propagator 
$\mathcal{G}_0^{-1} (i \omega_n) = i \omega_n+\mu - \Delta(i \omega_n)$,
the ``dynamical mean field''. The latter has to be determined
through a self-consistency condition, expressing the translational
invariance of the solid and thus the equivalence of different
atomic sites. For further details we refer the interested reader
to the many excellent reviews about DMFT \cite{georges, vollkot, 
pruschke_review, held-psik, biermann_ldadmft}.
In the present context, we restrict ourselves to a discussion of
how the above construction is modified when dynamical interactions
are taken into account. What is the relevant impurity model
if we want to solve a lattice model with purely local but dynamical
effective Hubbard interactions?
The answer is a straightforward generalisation to frequency-dependent
$\mathcal{U}(\omega)$ of the above action, where $S_{int}[\mathcal{U}]$
above replaces the static Hubbard interaction term:
\begin{eqnarray}
S = 
- \int \int d\tau d\tau^{\prime} \sum_{\sigma} c^{\dagger}(\tau^{\prime})
(\delta(\tau - \tau^{\prime}) \partial_{\tau} -
\Delta(\tau - \tau^{\prime})) c(\tau)
+ \int \int d\tau d\tau^{\prime} \mathcal{U}(\tau- \tau^{\prime})
n(\tau) n(\tau^{\prime})
\end{eqnarray}
where $n(\tau)= n_{\uparrow}(\tau)+ n_{\downarrow}(\tau)$.

An extremely efficient scheme for the solution of this
problem, suitable in the antiadiabatic regime,
is the recently introduced \cite{PhysRevB.85.035115} ``Boson
factor ansatz'' (BFA). It approximates the local 
Green's function of the dynamical impurity model as
follows:
\begin{equation}
G(\tau )= - \langle \mathcal{T} c(\tau) c^{\dagger}(0) \rangle
=
\left( \frac{G(\tau )}{G_{stat}(\tau )}\right) 
G_{stat}(\tau )
\sim
\left( \frac{G(\tau )}{G_{stat}(\tau )}\right) {\bigg|}_{\Delta=0}
G_{stat}(\tau )
\label{factorisation-approximation}
\end{equation}%
where G$_{stat}$ is the Green's function of a fully interacting
impurity model with {\it purely static interaction} U=$%
U(\omega =0)$, and the first factor is approximated by its
value for vanishing bath hybridization $\Delta$ \cite{PhysRevB.85.035115}.
In this case, it can be analytically evaluated in terms of 
the frequency-dependent interaction:
\begin{equation}
B(\tau) =
\left( \frac{G(\tau )}{G_{stat}(\tau )} \right) {\bigg|}_{\Delta=0}
= e^{- \int_0^{\infty} \frac{d\omega}{\pi}
\frac{\mathrm{Im} U(\omega)}{\omega^2}( K_{\tau}(\omega) - K_{0}(\omega))}
\label{factorisation-approximation} 
\end{equation}%
with $K_{\tau}(\omega)=\frac{exp(-\tau \omega) + exp(-(\beta- \tau) \omega)}
{1 - exp(-\beta \omega)}$.
In the regime that we are interested in, namely 
when the plasma frequency that characterises the variation of
$U$ from the partially screened to the bare value, is
typically several times the bandwidth, this is an excellent approximation, as
was checked by benchmarks against direct Monte Carlo calculations in
Ref.~\cite{PhysRevB.85.035115}. 
The reason can be understood when considering the solution of the
dynamical local model in the {\it dynamical atomic limit}
$\Delta=0$, that is,
when there are no hopping processes possible between the impurity
site and the bath. In this case the BFA trivially yields the exact
the solution, and the factorisation can be understood as a factorisation
into a Green's function determined by the static Fourier component
of $\mathcal{U}$ only and the exponential factor $B$ which only depends
on the non-zero frequency components of $\mathcal{U}$.
The former fully determines the low-energy spectral function of the
problem, while the latter is responsible for generating high-energy
replicae of the low-energy spectrum.
For finite bath hybridisation, the approximation consists 
in assuming that the factorisation still holds and that the
finite bath hybridisation modifies only the low-energy
static-$U$ Green's function, leaving the general structure of the
plasmon replicae generation untouched. The approximation thus
relies on the energy scale separation between low-energy processes
and plasmon energy; it becomes trivially exact not only in the
atomic limit but also in the static limit, given by small
electron-boson couplings or large plasmon energy.

The BFA lends a precise mathematical meaning to the physical
discussion of the generation of plasmon replicae. Indeed, the
factorisation of the Green's function corresponds in frequency
space to a convolution of the spectral representations of the
low-energy Green's function $G_{static}$ and the bosonic factor $B$.
In terms of the spectral
function $A_{stat}(\omega)$ of the static Green's function $G_{stat}(\omega)$
and the (bosonic) spectral function $B(\epsilon)$ of the bosonic
factor $B(\tau)$ defined above
the spectral function $A(\omega)$ of the full Green's function
$G(\tau)$ reads:
\begin{equation}
A(\omega)=\int_{-\infty}^\infty \!\!\! d\epsilon ~ B(\epsilon) 
\frac{1+e^{-\beta\omega}}{(1 + e^{-\beta(\epsilon-\omega)})(1 - e^{-\beta\epsilon})} 
A_{stat}(\omega-\epsilon).
\label{spectral_conv}
\end{equation}
In the case of a single mode of frequency $\omega_0$, the bosonic
spectral function consists of sharp peaks at energies given by
that frequency, and the convolution generates replicae of the
spectral function $A_{stat}(\omega)$ of the static part.
Due to the overall normalisation of the spectral function,
the appearance of replicae satellites is necessarily accompanied
by a transfer of spectral weight to high-energies. This 
mechanism induces a corresponding loss of spectral weight in the
low-energy part of the spectral function. Indeed, it can be
shown \cite{casula_effmodel}
that the spectral weight corresponding to the low-energy
part as defined by a projection on zero boson states is 
reduced by the factor
\begin{eqnarray}
Z_B & = & \exp \left( - 1/\pi  \int_0^\infty \!\!\! d \nu 
  ~  \textrm{Im} \mathcal{U}(\nu) /\nu^2 \right).
\label{Z_B}
\end{eqnarray}
Estimates of $Z_B$ for typical transition metal oxides vary
between $0.5$ and $0.9$, depending on the energy scale of
the plasma frequency and the efficiency of screening
(as measured e.g. by the difference between bare Coulomb
interaction 
$\langle |\frac{1}{|r - r^{\prime}|}| \rangle = \mathcal{U}(\omega=\infty)$
and the static value $\mathcal{U}(\omega=0)$).

\begin{figure}[t]%
\centering
\includegraphics[scale=0.69]{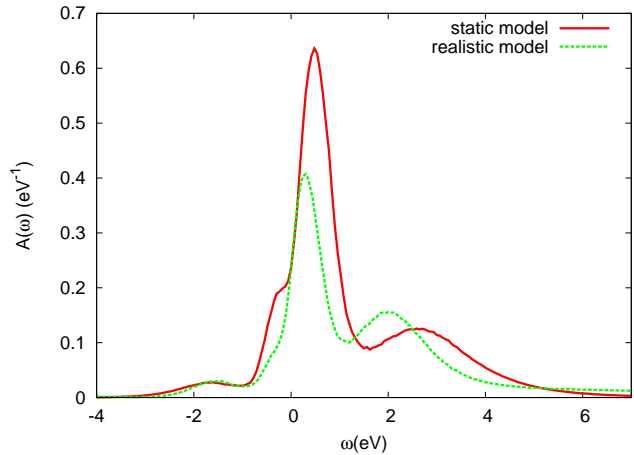}
\caption{Spectral function of a low-energy (t$_{2g}$-only) Hamiltonian
within ``LDA+$\mathcal{U}(\omega)$+DMFTT'' as compared to a
standard LDA+DMFT calculation with static interactions, see text.
From \cite{PhysRevB.85.035115}.
}%
\label{svo-dynU}%
\end{figure}

We reproduce in Fig.\ref{svo-dynU} the low-energy spectral function
of an ``LDA+$\mathcal{U}(\omega)$+DMFT'' calculation for the
d$^1$ ternary transition metal perovskite 
SrVO$_3$, demonstrating the reduction of spectral weight
compared to a static-U calculation \cite{casula_effmodel}. It should be noted 
however, that the calculation included the t2g states
only. We will show below that the contribution of unoccupied
e$_g$ states dominates at energies as low as $\sim 2.5$ eV.
Also, non-local self-energy effects stemming from screened
exchange interactions are non-negligible in this compound
and alter quite considerably the unoccupied
part of the t2g spectrum. We will come back to this point
below, within the discussion of fully dynamical GW+DMFT calculations
for SrVO$_3$ \cite{jmt_svo, jmt_svo2}.

\section{The Example of the Iron Pnictide BaFe$_2$As$_2$}

As a non-trivial example for the generalised ``LDA+$\mathcal{U}(\omega)$+DMFT"
approach, we review calculations on the iron pnictide compound BaFe$_2$As$_2$.
This materials is the prototypical compound of the so-called ``122-family''
of iron pnictide superconductors.
It exhibits superconductivity under pressure \cite{alireza,kimber} or hole- as well as electron-doping \cite{rotter,sefat}. Many  experimental probes  including angle-resolved and angle-integrated photoemission spectroscopy 
\cite{liu,brouet,Ding08,Borisenko09,Fink10, Zhang11, Mansart11},  optics and transport, Raman and  neutron scattering and nuclear magnetic resonance have been employed to characterise the electronic properties \cite{Wen11}. 
Experimental estimates of the (doping-dependent) mass enhancements vary substantially with doping; literature values range from about $1.4$ \cite{Yi09} to $5$, at least for the orbital pointing towards the As-sites \cite{Mansart11}.   The orbital character of the Fermi surface pockets are still subject to debate, but there seems to emerge a consensus about stronger correlation effects for  holes than electrons.

\begin{figure}[t]%
\centering
\includegraphics[width=0.69\textwidth]{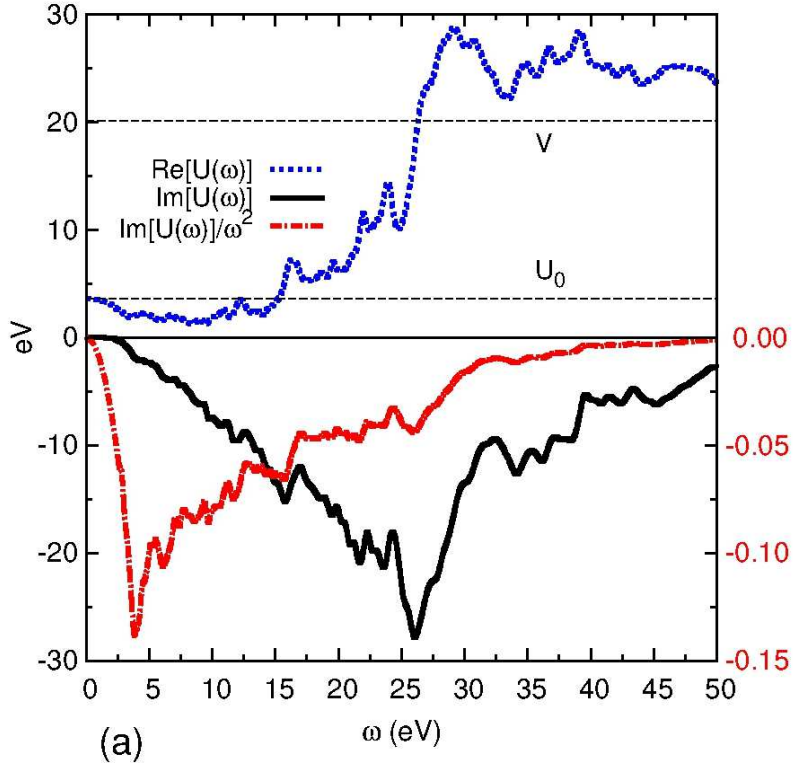}%
\caption{Frequency-dependent Hubbard interaction
$\mathcal{U}(\omega)$ for BaFe$_2$As$_2$: real and imaginary
parts as well as the mode distribution 
function $Im \mathcal{U}(\omega)/\omega^2$. From \cite{werner_Ba122}.
}
\label{ufreq}%
\end{figure}

\begin{figure}[t]%
\centering
\includegraphics[width=0.69\textwidth]{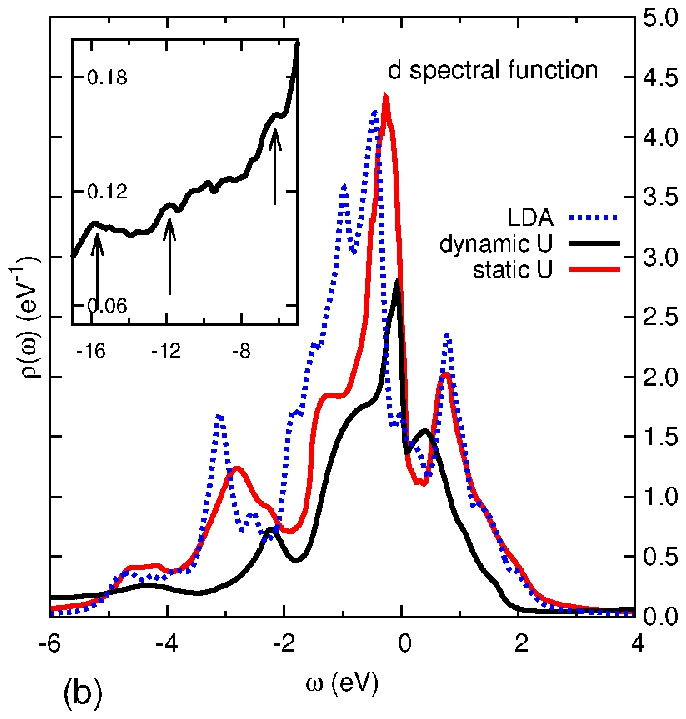}
\caption{Spectral function (3d-states only) of
 BaFe$_2$As$_2$ within LDA+$\mathcal{U}(\omega)$+DMFT, in 
comparison to standard LDA+DMFT and LDA,
see
text. From \cite{werner_Ba122}.
}
\label{spectra}%
\end{figure}

The constrained RPA result for the average intra-orbital Coulomb repulsion $\mathcal{U}$  of BaFe$_2$As$_2$ is shown in Fig.~\ref{ufreq}a.  
Here, $\mathcal{U}(\omega)$ represents a {\it partially screened} Coulomb interaction for the Fe-$d$ states, which accounts for screening
by all degrees of freedom except the Fe-$d$ states themselves. The real part ranges from the static value $U_0\equiv \text{Re}U(\omega=0)=3.6$~eV  to the bare interaction $V$ of about $20$~eV at large $\omega$.
In this case, screening does not arise from a single well-defined plasmon excitation. Instead, $\text{Im}\mathcal{U}(\omega)$ is characterized by a broad structure beginning from  a peak at  $\sim 26$~eV and  extending down to a few eV, implying that any plasmon excitations overlap strongly with the one-particle excitations. We note moreover that the term ``plasmon'' is used here
a bit abusively for any bosonic excitation mode that screens the
Coulomb interactions, regardless of the precise nature (plasmon,
particle-hole excitations or any other many-body satellite feature).

In a standard DFT+DMFT calculation without dynamical effects, 
the relatively small value 
of the  interaction $U_0$ would result in a rather weakly correlated picture. This is demonstrated in Fig.~\ref{spectra}, which presents  the Fe-$d$ spectral  function obtained by using 
the static $U_0$ in a standard DFT+DMFT calculation. Interaction effects  lead to a moderate renormalization of the Fe-$d$ states, with a mass enhancement of $1.6$.  Comparison to the DFT density of states 
shows that the peaks at  $-3$~eV and $1$~eV are weakly renormalized band states.  No Hubbard satellites or other correlation features appear, in agreement with previous studies \cite{skornyakov, Yin_pnictide}. 

The function $\frac{\text{Im}U(\omega )}{\omega ^{2}}$
is  plotted as the red dash-dotted line in Fig.~\ref{ufreq}. Besides a first peak at  $3.8$~eV, which comes from the rapid decay of $\text{Im}U(\omega )$ at small frequencies, there are prominent peaks at $6.1$~eV, $16$~eV and $26$~eV, as well as smaller features at $10$~eV and $12$~eV.  

The sharp low-energy peak in the $d$-electron spectral function results in weak, but well-defined satellites, as discussed above. The inset of Fig.~\ref{spectra} shows the high energy tail of the occupied part of the spectrum, with arrows marking the most prominent satellites at $-6.1$~eV, $-12$~eV and $-16$~eV. 
The observation of satellites at $-6.5$~eV and $-12$~eV was emphasized in the photoemission study of Ding and collaborators \cite{Ding08}. 
While Ref.~\cite{Jong09}   
confirms a hump in the $d$-electron spectral function around $-6.5$~eV, these authors suggest that the feature at $-12$~eV is an As-4$s$ line. Our calculation suggests that a $d$-feature, originating from the structure in the frequency dependent interaction, is superimposed to the As-4$s$ spectral contribution.  The $-16$~eV feature is probably not visible in experiments, because it overlaps with Ba-5p states,  while a satellite which we predict at $-3.8$~eV is masked by structures arising from $p$-$d$ hybridization.

A detailed analysis of the many-body self-energy
revealed a further interesting aspect of this compound,
namely a pronounced ``non-Fermi liquid'' (incoherent) regime
in the metallic phase near optimal doping, characterised by square-root
behavior of the self-energy. We refer the interested reader to the
original Ref. \cite{werner_Ba122}, where also implications for the doping 
and temperature 
dependence of the low energy electronic structure, in comparison with
angle-resolved photoemission were discussed.

\section{The combined ``GW+DMFT'' approach}

The solution of the DMFT equations
for a frequency-dependent (dynamical) Hubbard interaction, is also a
key step in the combined ``GW+DMFT'' method, as proposed in \cite{gwdmft}. 
Indeed, this scheme was proposed a few years ago, in order to
avoid the ad hoc nature of the Hubbard parameter and the double counting
inherent to conventional combinations of dynamical mean field theory with
the LDA. 
Moreover, the theory provides momentum dependence to quantities 
(such as the self-energy) that are local within pure DMFT. 

The starting point is Hedin's GW approximation 
(GWA)\cite{Hedin-1965, hedin}, 
in which the
self-energy of a quantum many-body system is obtained from a convolution 
(or product) of
the Green's function G with the screened Coulomb interaction 
$W=\epsilon^{-1}V$. 
The dielectric function $\epsilon $, which screens the bare Coulomb
potential $V$, is -- within a pure GW scheme -- obtained from the random
phase approximation. The GW+DMFT scheme, as proposed in 
\cite{gwdmft}, 
combines the first principles description of screening inherent in GW
methods with the non-perturbative nature of DMFT, where local quantities
such as the local Green's function are calculated to all orders 
in the interaction from an
effective reference system 
(\textquotedblleft impurity model\textquotedblright )%
\footnote{%
The notion of locality refers to the use of a specific basis set of
atom-centered orbitals, such as muffin-tin orbitals, or atom-centered Wannier
functions.}. 
In DMFT, one imposes a self-consistency condition for the one-particle
Green's function, namely, that its on--site projection equals the impurity 
Green's function. In GW+DMFT, the self-consistency requirement is
generalized to encompass two-particle quantities as well, namely, the local
projection of the screened interaction is required to equal the  
impurity screened interaction. This 
in principle promotes the Hubbard U from an adjustable parameter in
DMFT techniques to a self-consistent auxiliary function that incorporates
long-range screening effects in an {\it ab initio} fashion. 
Indeed, as already alluded to above and further elaborated upon
in Sect.~7, not only higher energy degrees
of freedom can be downfolded into an effective dynamical interaction,
but one can also aim at incorporating non-local screening effects
into an effective dynamical $\mathcal{U}(\omega)$.
The theory is then free of any Hubbard {\it parameter}, and the
interactions are directly determined from the full long-range
Coulomb interactions in the continuum%
\footnote{Within the terminology of Hedin's equations, this
means in particular that screening is assessed beyond the
random phase approximation. Fully self-consistent GW+DMFT
are rare, and have so far been performed only in the case of
an extended Hubbard model \cite{PhysRevLett.92.196402,
ayral, ayral-prb} and for the system of adatoms on
semiconductor surfaces \cite{hansmann} discussed below.
In these studies, it was in particular shown that in the
regime close to the metal-insulator transition, the RPA
yields a poor estimate for screening which is strongly
suppressed by correlation effects, see \cite{ayral}.}.

From a formal point of view, the GW+DMFT method, as introduced in \cite{gwdmft}, 
corresponds to a specific 
approximation to the correlation 
part of the free energy of a solid, expressed as a functional of the Green's
function G and the screened Coulomb interaction W: the non-local part is
taken to be the first order term in $W$, while the local part is calculated
from a local impurity model as in (extended) dynamical mean field theory.
This leads to a set of self-consistent equations for the Green's function 
$G$, 
the screened Coulomb interaction $W$, the self-energy $\Sigma $ and the
polarization $P$ \cite{gwdmft_proc2,gwdmft_proc1} 
(which are reviewed in Appendix C).
Specifically, the self-energy is obtained as $\Sigma
=\Sigma _{local}+\Sigma _{non-local}^{GW}$, where the local part 
$\Sigma_{local}$ is derived from the impurity model. 
In practice, however, the calculation of a self-energy for
(rather delocalized) s- or p-orbitals has never been performed 
within DMFT, and it appears to be more physical to approximate 
this part also by a GW-like expression. 
For these reasons Ref. \cite{jmt_svo, jmt_svo2} proposed a practical scheme,
in which only the local part of the self-energy of the \textquotedblleft
correlated\textquotedblright\ orbitals is calculated from the impurity model
and all other local and non-local components are
approximated by their first order expressions in $W$. 

\section{From ``LDA+$\mathcal{U}(\omega)$+DMFT''
to ``GW+DMFT'': the Example of SrVO$_3$}

The very first dynamical (albeit not yet fully self-consistent) implementation 
of ``GW+DMFT'' was achieved in 2012, and applied to the ternary transition
metal oxide SrVO$_3$ \cite{jmt_svo, jmt_svo2}. 
In this section, we review the results of these
calculations, before entering more in detail into questions of the 
formalism and finally describing a -- fully self-consistent -- implementation
in the single-orbital case and its application to surface systems 
(see Section 8). 

Our target material, SrVO$_{3}$, has been thoroughly studied, both,
experimentally and theoretically. It cristallizes in the cubic
perovskite structure, splitting the V-d states into a threefold degenerate $%
t_{2g}$ manifold, filled with one electron per V, and an empty $e_{g}$
doublet. It has been characterized as a correlated metal with a
quasiparticle renormalization of about 0.6 \cite{PhysRevB.73.052508,PhysRevLett.93.156402,
PhysRevB.80.235104}%
, and a photoemission (Hubbard-) satellite at around -1.6 eV binding
energy \cite{PhysRevB.52.13711}. Inverse photoemission has located the
electron addition $d^1 \rightarrow d^2$ peak at an energy of about
2.7 eV \cite{PhysRevB.52.13711}.

\begin{figure}[htb]
\centering
\includegraphics[clip=true,
,angle=-90,width=.69\textwidth]{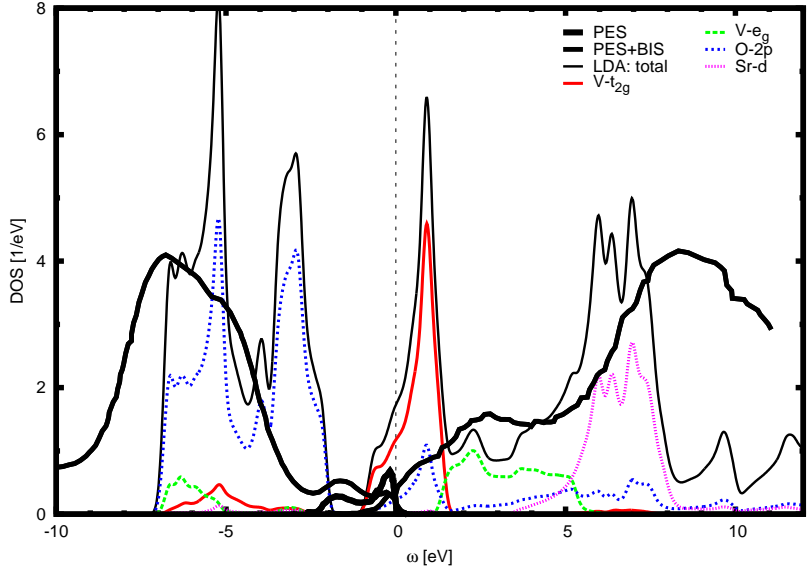}
\caption[bla]{
Orbital-resolved LDA density of states, in comparison to experimental
spectra from photoemission (PES)\cite{PhysRevB.52.13711,PhysRevLett.93.156402} and inverse photoemission (BIS)\cite{PhysRevLett.93.156402}.
Adapted from \cite{jmt_svo}.
}
\label{fig:lda-svo}
\end{figure}

\begin{figure}[h]
\centerline{\includegraphics[clip=true,
angle=-90,width=.65\textwidth]{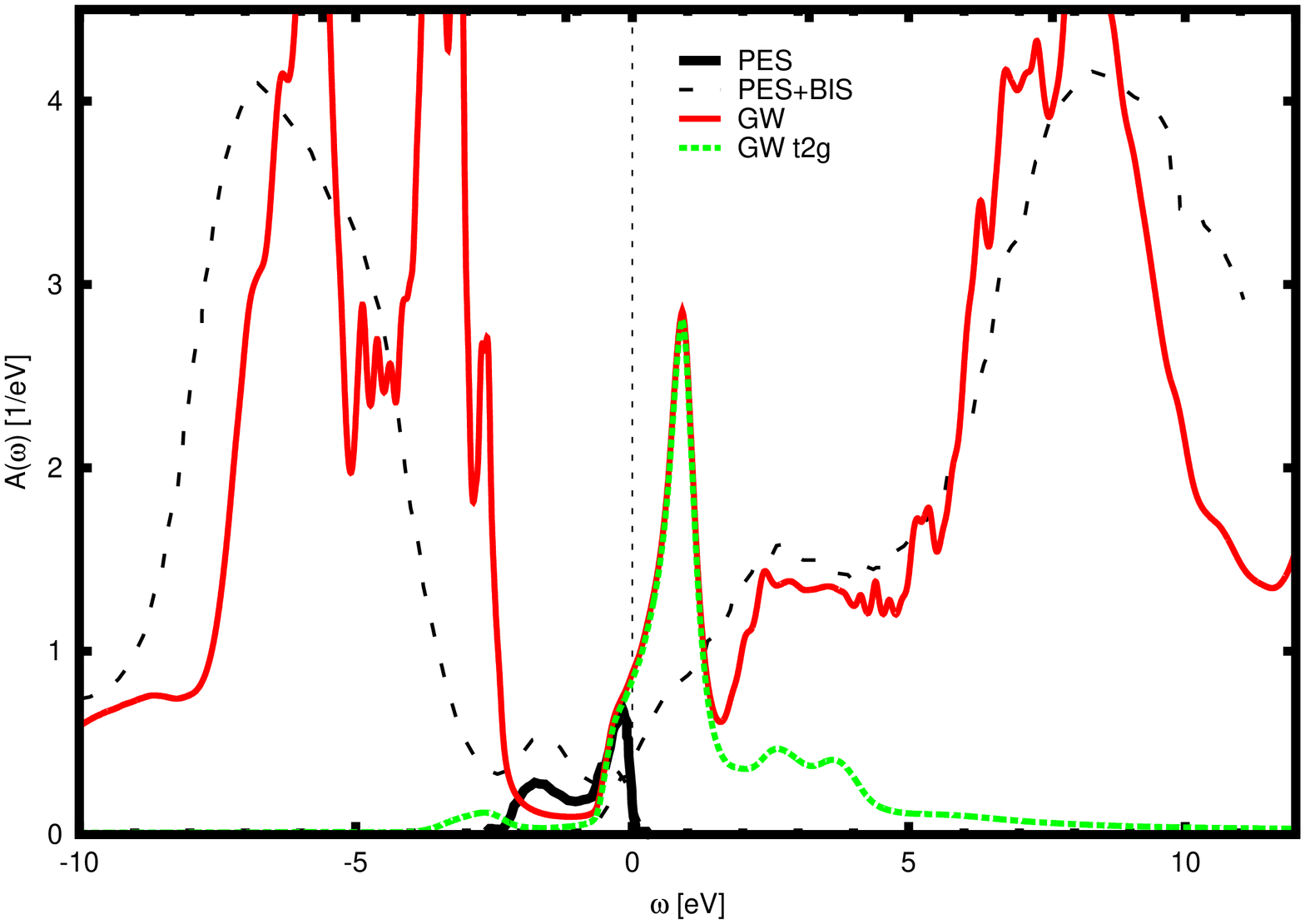}
}
\caption[bla]{
Spectral function from GW calculations 
in comparison to experiments.
The \t2g\ orbital contribution is resolved as dashed line.
The Fermi energy is set to zero.
Adapted from \cite{jmt_svo2} (see also \cite{jmt_svo}).}
\label{fig:gw-svo}
\end{figure}

Figure (\ref{fig:lda-svo}) summarizes the LDA electronic structure:
the O-p states disperse between -2 and -7 eV, separated from the
\t2g states whose bandwidth extends from -1 eV to 1.5 eV.
While the \t2g and e$_{g}$ bands are well separated at every given k-point,
the partial DOS slightly overlap, and the e$_{g}$ states display a pronounced
peak at 2.3 eV. Finally, peaks stemming from the Sr-d states are
located at  6.1 eV and 7.1 eV.
We have superimposed to the LDA DOS the experimental PES and BIS 
curves taken from \cite{PhysRevB.52.13711,PhysRevLett.93.156402}. The comparison reveals the main
effects of electronic correlation in this material: as expected on 
quite general grounds, LDA locates the filled O-p states at too high 
and the empty Sr-d manifold at too low energies.
The \t2g manifold undergoes a strong quasi-particle
renormalization and a concomitant shift of spectral weight to the lower
Hubbard band, both of which are effects beyond the one-particle
picture.
The GW approximation (see the spectral function in Fig.(\ref{fig:gw-svo}))
increases the O-p to Sr-d distance, placing both manifolds at
energies nearly in agreement with experiment \footnote{%
The persisting slight underestimation is expected, since
the polarization function that determines $W$ is calculated
using the bare t$_{2g}$ bands, whose itinerant character
is largely overestimated by the LDA.
}.
Most interestingly, however, a peak at 2.6 eV emerges from the
d-manifold, which we find to be of e$_{g}$ character. Indeed, the
GW approximation enhances the t$_{2g}$-e$_g$ 
splitting and
places the maximum of the e$_{g}$ spectral weight at the location of
the experimentally observed $d^1 \rightarrow d^2$ addition peak.
The panel also displays the GW partial \t2g contribution.
These states show two interesting features: their width is
narrowed from the LDA value by about 0.5 eV and satellite 
structures appear below and above the main quasiparticle peak
at $\pm$3~eV, as well as above 15eV. The origin of these features
was analysed in \cite{jmt_svo, jmt_svo2}, where it was argued that the peak
at 15eV corresponds to the physical plasmon discussed in
Section 2, while the lower energy satellites are spurious
features due to the perturbative nature of the GW approximation.

In the implementation of Ref. \cite{jmt_svo, jmt_svo2} 
the GW+DMFT equations were
solved self-consistently only at the DMFT level, that is for a fixed
screened interaction (corresponding to the cRPA one) and fixed non-local
GW self-energies. However, the fully dynamical interactions were
retained, and the GW+DMFT equations solved within the BFA 
\cite{PhysRevB.85.035115} reviewed above.

\newpage

\

\begin{figure}[h]
\includegraphics[clip=true, 
angle=-90,width=.65\textwidth]{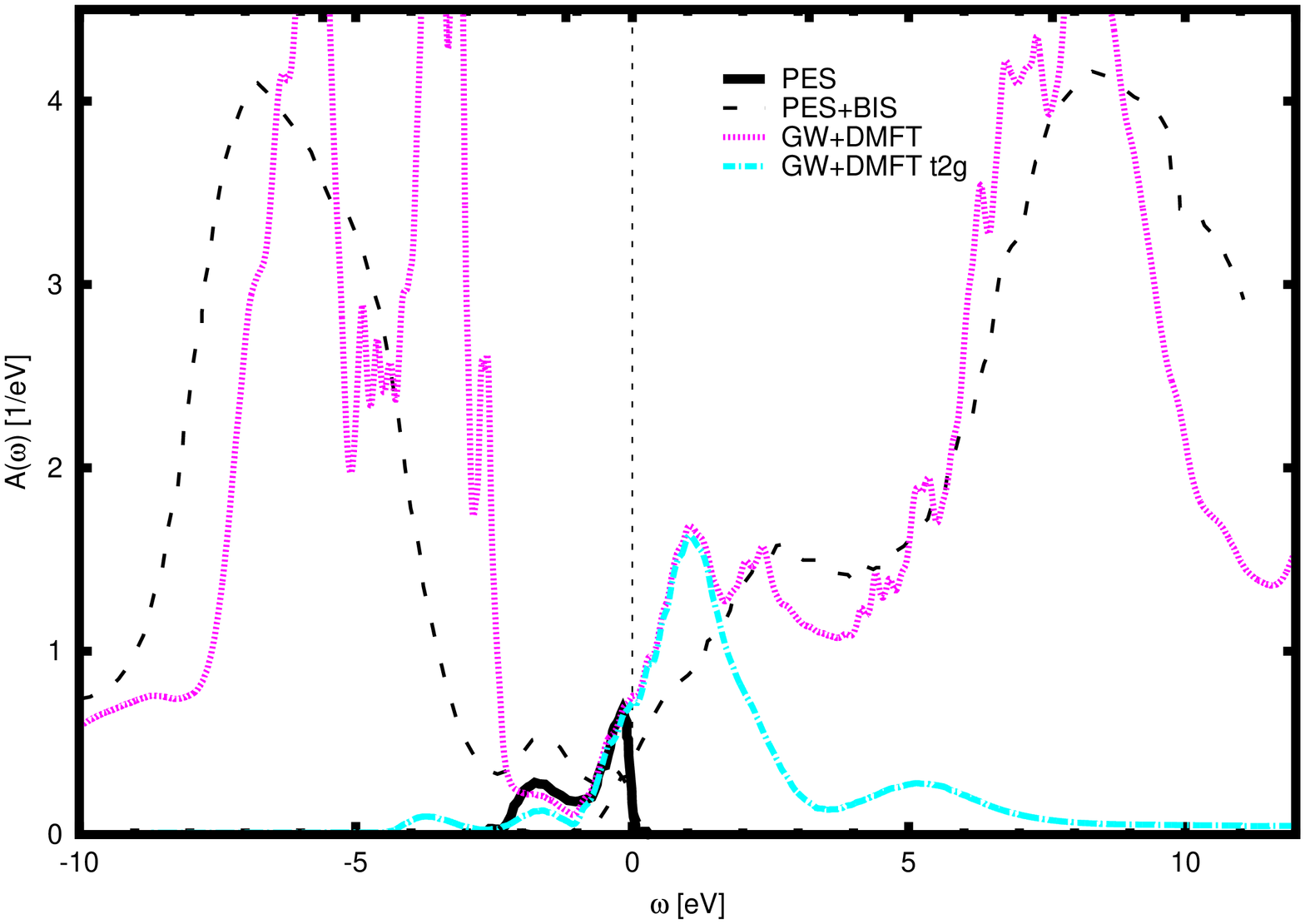}
\includegraphics[clip=true,
angle=-90,width=.35\textwidth]{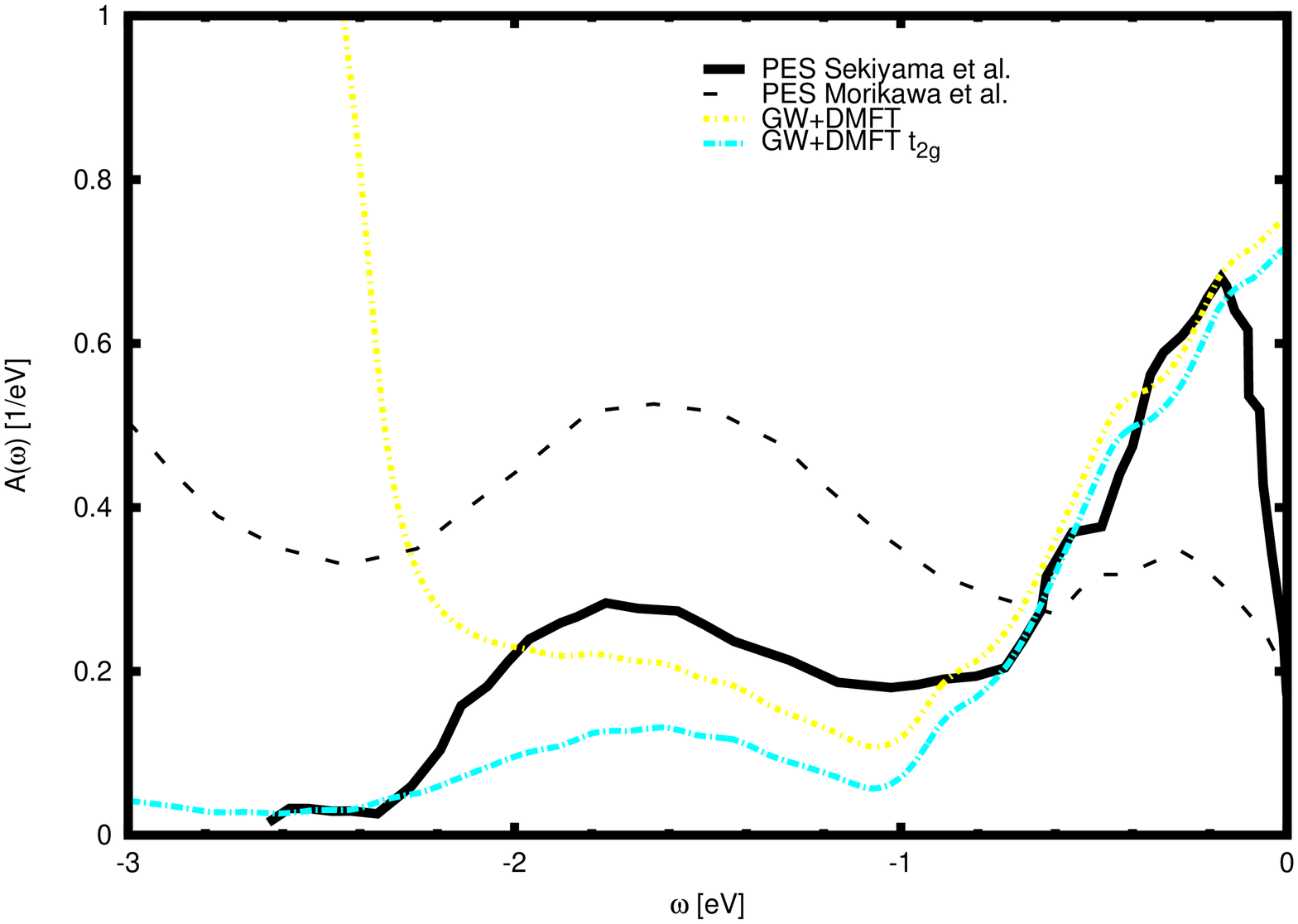}
\caption[bla]{
Spectral function from GW+DMFT in comparison to experiments.
The \t2g\ orbital contribution is resolved as dashed line. 
The Fermi energy is set to zero.
Adapted from \cite{jmt_svo2} (see also \cite{jmt_svo}).}
\label{fig:gw-svo2}
\end{figure}

\begin{figure}[t!h]
\centering
\includegraphics[clip=true,trim=5 0 40 0, angle=-90,width=0.69\textwidth]
{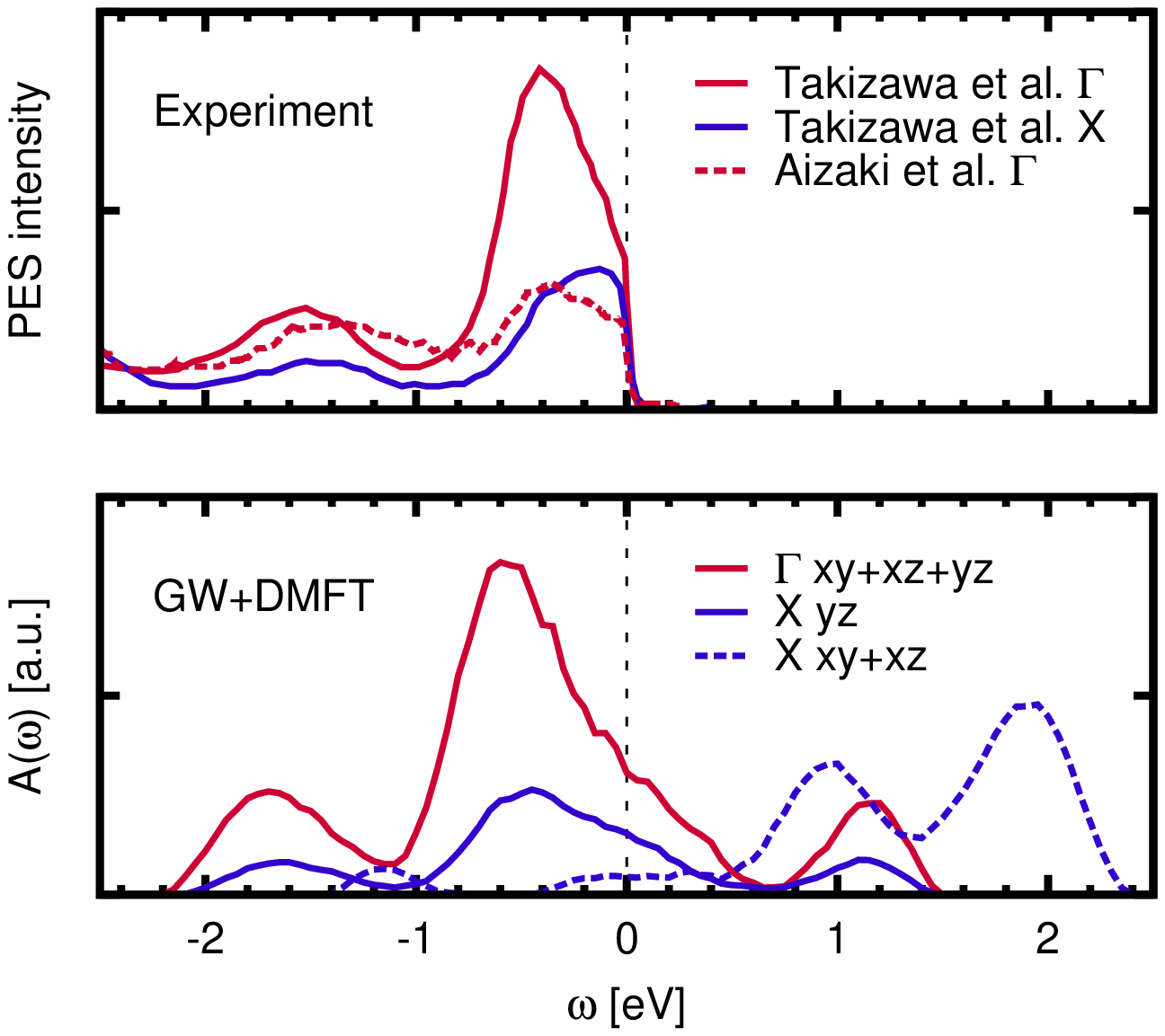}
\caption[bla]{Momentum-resolved \t2g\ spectral function  
from GW+DMFT, compared to experimental angle resolved photoemission
spectra of Refs.~\cite{aizaki_svo}
and \cite{PhysRevB.80.235104}. 
Adapted from Ref.~\cite{jmt_svo}.}
\label{fig:arpes} 
\end{figure}

The results for the spectrum of SrVO$_{3}$ are plotted in Fig \ref{fig:gw-svo},
which displays the local spectral function. 
In Fig~\ref{fig:arpes} we compare the momentum resolved GW+DMFT spectral function to
pure GW results as well as to the most
recent angle resolved photoemission experiments (ARPES)\cite{aizaki_svo}. 
The low-energy part of the spectral function is dominated by
the \t2g contribution, which is profoundly modified within
GW+DMFT compared to the pure GW spectrum. A renormalized
quasi-particle band disperses around the Fermi level, with 
e.g. at $\Gamma$ a sharp peak in the occupied part of the
spectrum at about -0.5 eV binding energy, corresponding to
a strong renormalisation of the LDA Kohn-Sham state that 
has, at this momentum, an energy of -1 eV.
At the X-point, the three \t2g bands are no longer degenerate,
and surprisingly weakly renormalized xz/yz states are observed
at 0.9 eV, while the yz band is located at nearly the
same energy as at the $\Gamma$ point. This is in excellent agreement 
with experiments:
In photoemission spectroscopy a renormalized 
quasi-particle band structure is dispersing between the 
$\Gamma$ and X points. At binding energies of -1.6 eV a weakly dispersive
Hubbard band forms, whose intensity varies significantly as a function 
of momentum \cite{PhysRevB.80.235104}. 
In the GW+DMFT spectral function, the Hubbard band is observed
at about -1.6 eV, and its k-dependent intensity
variation is indeed quite strong. As anticipated from the total
spectral function discussed above, the GW spectrum also displays
a satellite structure, which is however of plasmonic origin, 
arising from the structure in $\mathrm{Im}  W$.
This is a well-known failure of the GWA, which has been analyzed 
in detail in 
\cite{0953-8984-11-42-201}:
the simple form of the GW self-energy
is not able to encode multiple satellites, and the single
plasmon is then located at a too high energy.
The lower Hubbard band is absent in GW, as expected.

As discussed above, the low-energy features in $W$ are
absent from the dynamical Hubbard interaction entering the combined 
GW and dynamical calculation, consistently with the fact that
the full GW+DMFT calculation does not show spurious features
at 3 eV as does the GW.

The overall picture of the GW+DMFT spectra
results in an occupied band structure
that resembles closely the dynamical mean field
picture (see e.g. \cite{pavarini:176403})
though the lower Hubbard band is located
slightly closer to the Fermi level, at about -1.6 eV, 
in agreement with experiments. This improvement is an effect
of the relatively smaller zero-frequency $U$ value (3.6 eV) compared
to parameters commonly used in standard LDA+DMFT calculations
(around 4 to 5 eV). In the latter, larger $U$ values are required
to match the experimentally observed mass enhancements, thus
necessarily placing the lower Hubbard band at energies slightly too 
far away from the Fermi level.
The dynamical screening, included in our
calculations, results in additional spectral weight transfers 
\cite{PhysRevB.85.035115, casula_effmodel}, thus yielding at the same time a 
good description of mass enhancements and the Hubbard band.

In the unoccupied part of the spectrum non-local
self-energy effects are larger. 
Interestingly, our total spectral function,
right panel of Fig.\ref{fig:gw-svo}, does not display
a clearly separated Hubbard band. The reason is visible from the 
k-resolved spectra: the upper Hubbard band is located at around
2 eV, as expected from the location of the lower Hubbard band and
the fact that their separation is roughly given by the 
zero-frequency value of U (=3.6 eV).
The peak around 2.7 eV that appears in the inverse photoemission
spectrum \cite{PhysRevB.52.13711} -- commonly interpreted as the
upper Hubbard band of \t2g character in the DMFT literature -- 
arises from e$_{g}$ states located in this energy range.
The non-local
self-energy effects lead, in the unoccupied part of the
spectrum, to overlapping features from different k-points and
the impression of an overall smearing out of the
total spectral function.
From the k-resolved spectra, it is also clear that while
non-local self-energy effects stemming from the GW part
have little influence on the occupied part of the spectrum,
they widen the bands in the unoccupied part
substantially. A rough estimate of the various renormalisation
effects on the overall bandwidth leads to a -- at first sight
-- astonishing conclusion: effects of the dynamical tail of
the Hubbard $\mathcal{U}$ have been estimated to roughly
lead to a band renormalisation of $Z_B \sim 0.7$ \cite{casula_effmodel},
and the renormalisation due to the static part $\mathcal{U}(\omega=0)$
still adds to this. Nevertheless, the final position of the
empty quasi-particle bands after the GW+DMFT calculation
nearly coincides with the initial LDA energies. This gives
an order of magnitude for the substantial widening of the
band induced by 
nonlocal effects. The picture is consistent
with the observation of Refs.~\cite{PhysRevB.87.115110, jmt_qsGW}, that
a purely local GW calculation in fact leads to much stronger
renormalisation effects than the full GW calculation, and the
band structure within the latter results from subtle
cancelations of band narrowing
due to the local self-energy and widening due to its
non-local parts.
One could thus be tempted to conclude that renormalisation
effects due to local dynamical interactions and widening
due to non-local self-energies cancel, giving new justifications
to combined LDA++ schemes with static interactions.
There are several reasons why this conclusion would be too quick:
First, the widening effects rather selectively act on the
unoccupied band structure, since the exchange-correlation
potential of LDA is a much better approximation to the
many-body self-energy for occupied states than for empty ones.
Second, the renormalisation effect due to dynamical
interactions goes hand in hand with a spectral weight
loss at low-energies. These are barely observable in photoemission
spectroscopy since spectra are generally not measured in absolute
units, and even then would matrix element effects make a
comparison of absolute normalisations intractable.
Probes that can assess absolute units, such as optical
spectroscopy, however, can be expected to be sensitive
to such shifts of spectral weight.

\section{Generalised  ``downfolding'' of long-range Coulomb
interactions:  the combined ``GW+DMFT'' scheme 
from the ``representability point of view''}

In calculations of the effective local Coulomb 
interactions within the constrained random phase approximation
the dynamical nature of the
interaction is generally supposed to stem only from the downfolding
of higher-energy degrees of freedom%
\footnote{See however the recent works aiming at the construction
of an effective local Hubbard interaction to be used within the
DMFT context from a cRPA scheme where $P_{low}$ is restricted
to represent local screening only \cite{vaugierPHD, nomura}}. 
It can then be directly
assessed by a cRPA construction of a dynamical lattice model.
The construction of the impurity model serves in this case
merely as a tool to solve the dynamical lattice problem, with a
fixed local dynamical interaction that is assumed to be the local
part of the cRPA one. The GW+DMFT formalism, on the other hand,
demonstrates that it is possible to adopt a more
general point of view, that also incorporates the contribution
of nonlocal interactions and nonlocal screening effects in
a solid, giving rise to an additional frequency dependence
of the effective local interaction. The Hubbard interaction
U in this case should no longer be interpreted as the local
part of the physical Coulomb interaction of a downfolded
model, but as an effective quantity that incorporates both, the
effects of screening by downfolding and by representing a
lattice model by a local model. This perspective goes beyond the
cRPA view of the problem, but can be consistently formulated
within extended DMFT \cite{sengupta, si, kajueter} or 
GW+DMFT.
Indeed, when implemented in a fully self-consistent fashion,
GW+DMFT provides a
prescription on how to calculate both the one-particle part of
the Hamiltonian and the effective Hubbard interactions of a correlated
material from first principles. The idea 
is to calculate the nonlocal part not only of the self-energy 
but also of the polarization to lowest order in the screened Coulomb 
interaction. These non-local quantities are then combined
with their local counterparts as calculated from a
dynamical impurity model. One thus represents two physical
quantities, namely, the local Green’s function G of the solid
and the local part of the fully screened Coulomb interaction
W by a local model, defined by some effective Weiss field
$\mathcal{G}_0$ and the auxiliary Coulomb interaction 
$\mathcal{U}(\omega)$. The latter
is constructed such that the solution of the impurity model
yields the local part of W. This is akin in spirit to other
theories in solid state physics, where a physical quantity is
represented by the self-consistent solution of an effective
auxiliary model, famous examples being density functional
theory or DMFT itself. In DFT, the physical density of a
system is represented by an auxiliary system in an effective
one-particle (Kohn-Sham) potential; in DMFT, the local lattice
Green’s function is constructed from an impurity model with
an effective Weiss field or, equivalently, a local self-energy.
The auxiliary quantities such as the Kohn-Sham potential of
DFT or the impurity self-energy acquire the role of Lagrange
multipliers fixing the density (in DFT) or the local Green’s
function (in DMFT) to their physical values. In extended
DMFT, a nonlocal interaction in the original Hamiltonian
gives rise to a dynamical impurity model representing the
physical quantities of the model, and it is the polarisation
that takes over the role of the corresponding Lagrange multiplier.
This carries through to the combined GW+DMFT scheme where
the non-local polarisation moreover adds to the frequency-dependence
of the effective local interaction.
The formalism of GW+DMFT, leading to a closed set of coupled
equations for the one- and two-particle quantities $G$, $W$,
the one- and two-body self-energies $\Sigma$ and $P$ and
the auxiliary quantities $\mathcal{G}_0$ and $\mathcal{U}(\omega)$
is reviewd in Appendix C.

\section{Exploring the Self-consistent ``GW+DMFT'' scheme:
Systems of Adatoms on Semiconductor Surfaces}

While the first dynamical implementation of GW+DMFT for a real
material -- the calculation for SrVO$_3$ reviewed above -- was 
not yet fully self-consistent, the full self-consistent cycle 
has in the meanwhile
been explored on the example of a different class of materials:
Systems of adatoms on semiconducting surfaces, such as Si(111):X 
with X=Sn, C, Si, Pb, present the advantage that their electronic
structure is determined by a low-energy Hilbert space spanned
by a single narrow half-filled surface band, which lies -- 
well-separated -- in the gap of Si.
Proposed early on \cite{tosatti74} to be good candidates for 
observing low dimensional correlated physics, these systems
have for a long time been considered to be realisations of the 
one-band Hubbard model on the triangular lattice, and have been
the subject of a variety of experimental
(\cite{uhrberg85,grehk93,weitering93,carpinelli96,carpinelli97,
weitering97,lay01,pignedoli04,upton05,modesti07,cardenas09,zhang10,
tournier11}) and theoretical \cite{kaxiras90,brommer92,santoro99,
hellberg99,aizawa99,profeta00,shi02,shi04,profeta05,profeta07,
schuwalow10,chaput11,li11} studies.
Experimentally, the so-called $\alpha$-phases show a remarkable 
variety of interesting physics including commensurate charge 
density wave (CDW) states \cite{carpinelli96,carpinelli97,lay01} 
and isostructural metal to insulator transitions (MIT)\cite{modesti07}. 

In Ref. \cite{hansmann}, a low-energy effective Hamiltonians 
for describing the surface state was derived \emph{ab initio} from 
density functional theory and the constrained random phase approximation 
(cRPA) scheme \cite{PhysRevB.70.195104} in the implementation of 
\cite{PhysRevB.86.165105} (see also the extension to surface systems 
in \cite{jpcm12}).
The LDA band structure is plotted
in Fig. 10, and the results for the
interaction parameters are given in
Table 1.
The local interactions vary from 1eV to 1.4eV, depending 
on the adatom species. 
This is not only much larger than the bandwidths
($\sim$ 0.5eV) confirming indeed the importance of correlation
effects for these systems. It is also somewhat larger than what
was previously assumed in model studies that treated the Hubbard
interaction as adjustable parameter in order to fit experiments.
Even more interestingly, however, long-range interactions are
large, with the interaction between electrons in
Wannier orbitals located on nearest neighbor atoms being about
$50 \%$ of their onsite counterparts. 
The relevant low-energy Hamiltonian is then a single-orbital
Hubbard Hamiltonian, extended to comprise the full tail of Coulomb
interactions, where the value is reduced by screening but not the
range. We stress that all parameters entering these Hamiltonians
are calculated from first principles.
Momentum-resolved spectral functions -- to be compared to
angle-resolved (inverse) photoemission spectra -- and the
two particle charge susceptibility were then calculated within the fully 
self-consistent GW+DMFT scheme applied to the lattice Hamiltonians.
To this end, the GW+DMFT implementation of \cite{ayral,ayral-prb} was 
generalised to the case of realistic Hamiltonians and long-range
interactions using an Ewald technique.

\begin{table}[t]
\caption{\label{uvaluestab} Values of the 
static (partially) screened interactions ($U_0=U(i\nu=0)$) for on- and 
intersite nearest neighbor (nn) interaction parameters
as determined from cRPA for use as {\it bare} interactions within
the Hilbert space of the single surface band. 
Also reported are the values of the static 
self-consistent $\mathcal{U}^{\rm GW+DMFT}_{\omega=0}$, incorporating
non-local screening processes through the GW+DMFT ``spatial
downfolding'' procedure. These values should be compared to 
the energy-scale of the Hubbard-gap in the spectral function, see text.}

\begin{tabular*}{\columnwidth}{@{}l*{15}{@{\extracolsep{0pt plus12pt}}l}}
\hline\\[-0.3cm]
 &C&Si&Sn&Pb& \\
\hline\\[-0.2cm]
bandwidth & & & & & [meV] \\
\hline\\[-0.2cm]
$U_0$&$1.4$&$1.1$&$1.0$&$0.9$&[eV]\\
$U_1$&$0.5$&$0.5$&$0.5$&$0.5$&[eV]\\
$U_2$&$0.28$&$0.28$&$0.28$&$0.28$&[eV]\\[-0.2cm]
$\vdots$& & & & \\
$U_n$& & &$U_1/r_a$& &\\[0.1cm]
\hline\\[-0.2cm]
$\mathcal{U}^{\rm GW+DMFT}_{\omega=0}$
&$1.3$&$0.94$&$0.84$&$0.67 ({\rm ins.})$&[eV]\\
 & & & &$0.54 ({\rm met.})$&[eV]
\end{tabular*}
\end{table}

\begin{figure}[t]%
\includegraphics[width=\columnwidth]{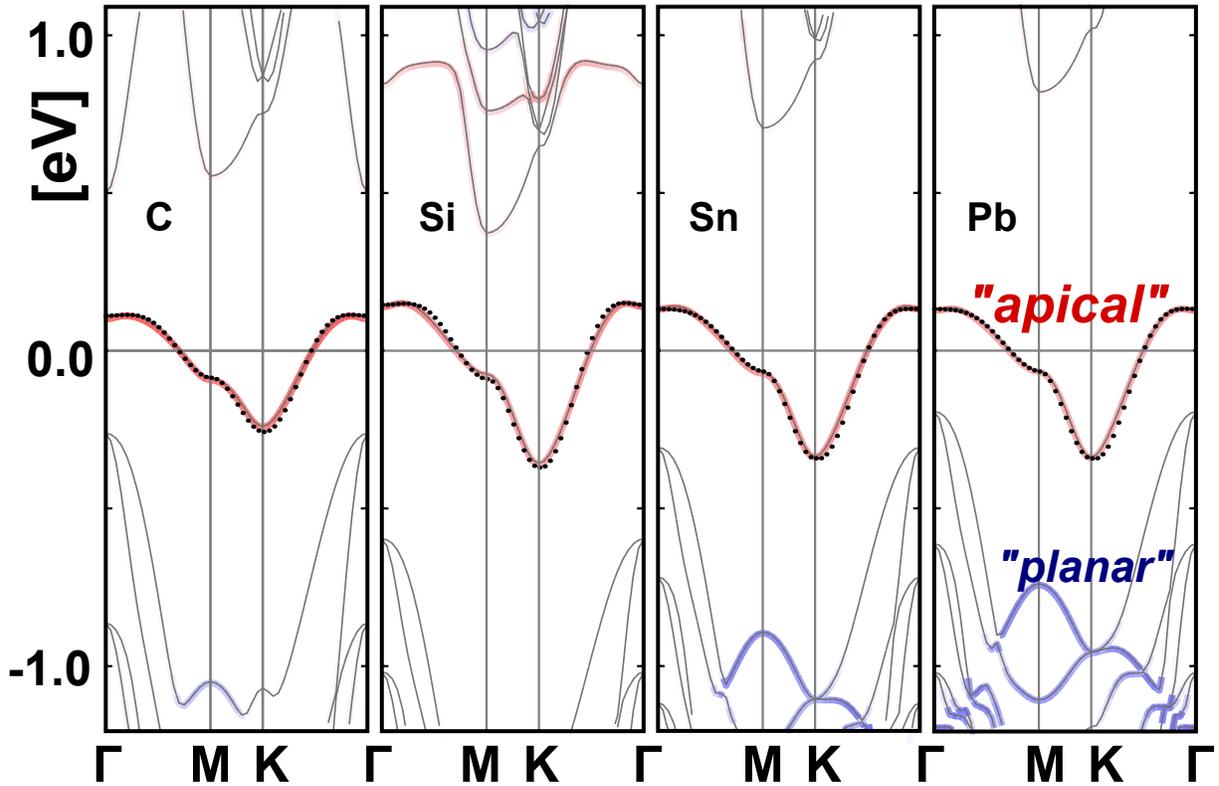}%
\caption{Bandstructures of the $\alpha$- $\sqrt{3}\times\sqrt{3}$ phases for Si(111):X with X=Sn, Si, C, Pb \cite{silicon}. The color of the bands denotes their respective orbital character. Red color indicates a $p_z$-like ``apical'' character, while the blue color denotes $p_{x,y}$-like (i.e. \emph{planar}) character.%
From Ref. \cite{hansmann}}%
\label{fatbands}%
\end{figure}

\begin{figure}[t]%
\centering
\includegraphics[width=\columnwidth]{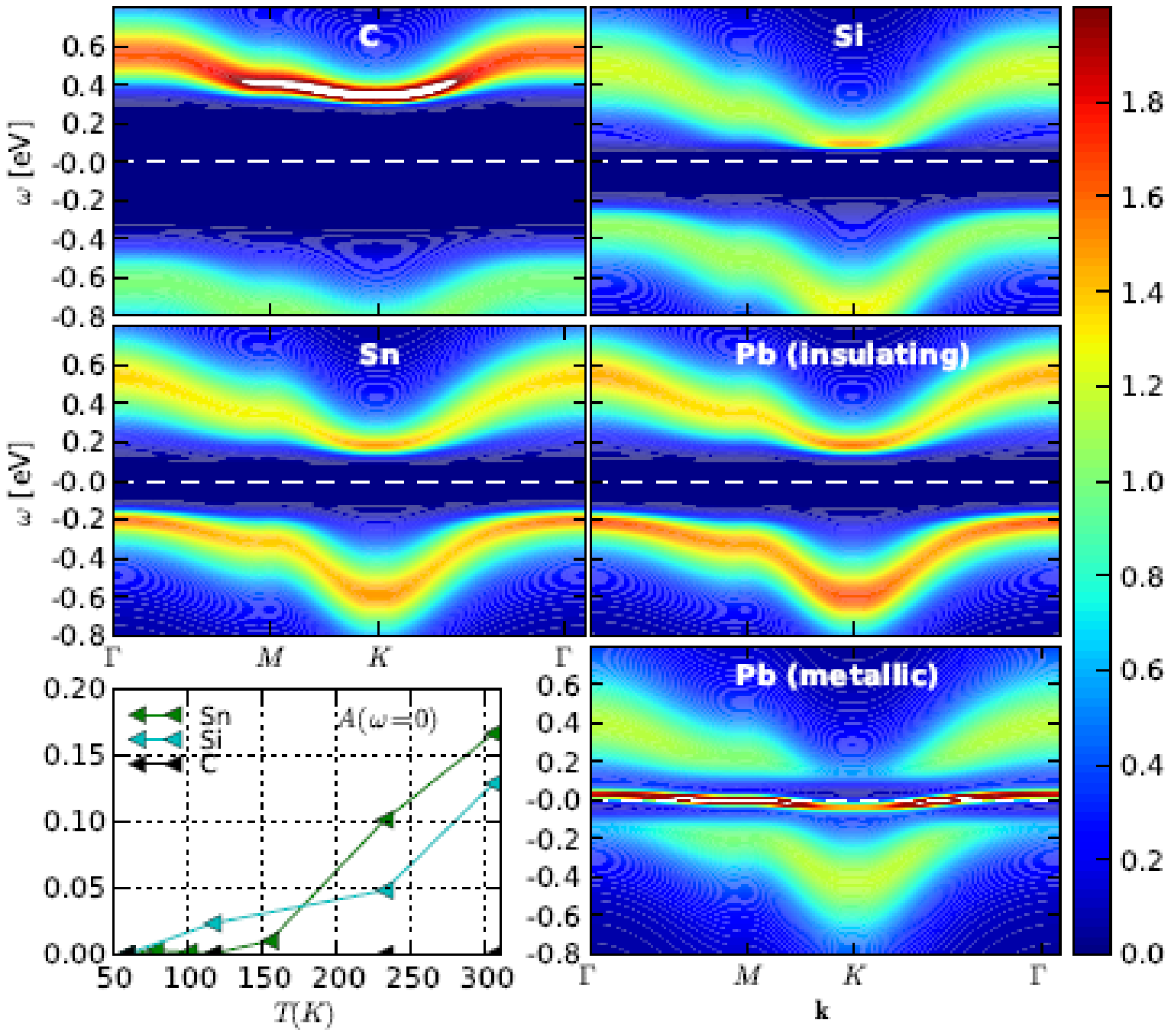}%
\caption{Momentum-resolved spectral function  at $T=116$K  of Si(111):X with X=Sn, Si, C, Pb obtained by analytical continuation of GW+DMFT imaginary time data. The Fermi energy is set to $\varepsilon_{\rm F}=0$ and indicated by the white dashed line.
From Ref. \cite{hansmann}%
}%
\label{fig02}%
\end{figure}

\begin{figure}[t]%
\centering
\includegraphics[width=\columnwidth]{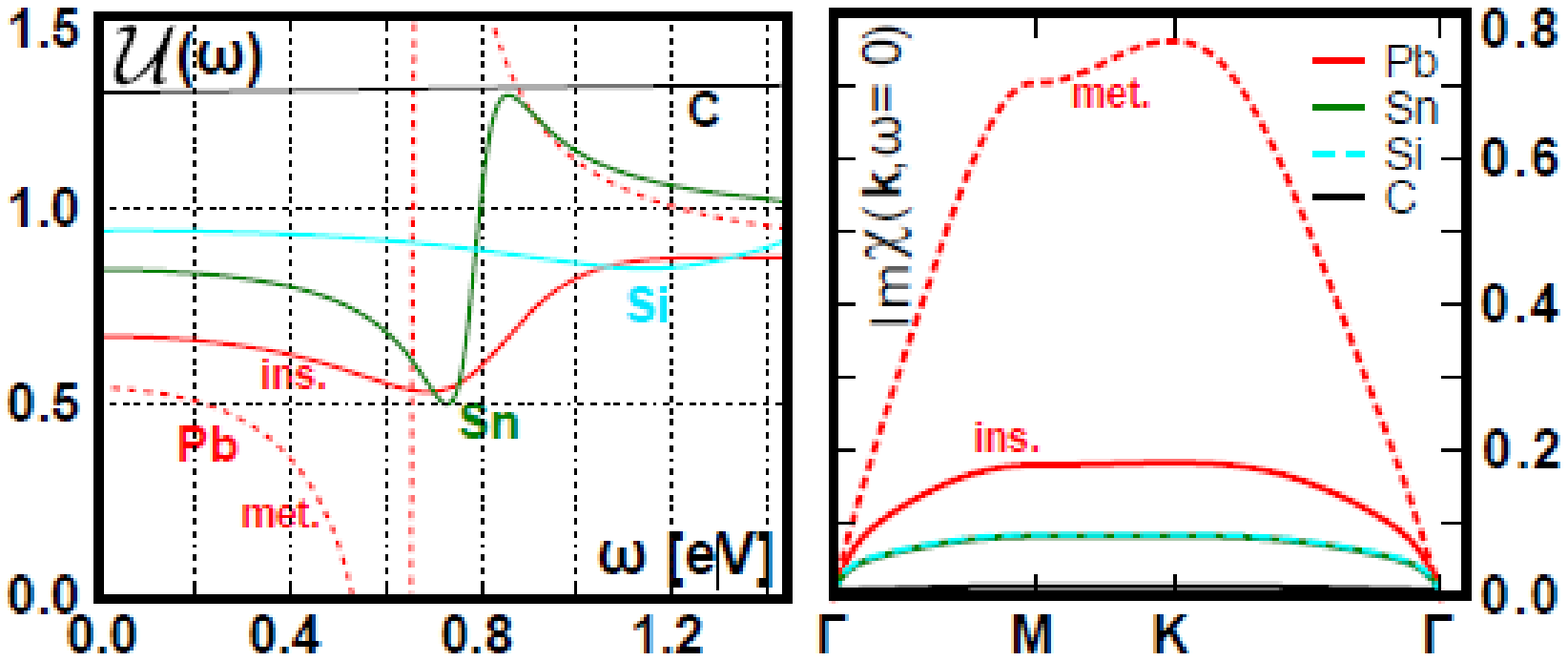}%
\caption{Left hand side: Frequency dependent $\mathcal{U}^{\rm GW+DMFT}(\omega)$ for all compounds including both, insulating and metallic cases for the Pb sytsem. Right hand side: Imaginargy part of the charge-charge susceptibility along the usual path in the Brillouin zone.
From Ref. \cite{hansmann}%
}%
\label{fig03}%
\end{figure}

Fig.~\ref{fig02} reproduces the momentum-resolved spectral functions 
for all compounds of the series:
As expected from the large onsite interactions compared to the bandwidth 
we obtain insulating spectra for all four compounds. Interestingly, however, 
for the Pb compound two stable solutions -- one metallic, one insulating --
were found at the temperature of our study ($T=116$K), stressing the
proximity of this system to the metal-insulator transition (and charge
order instabilities, as born out of an analysis of the charge-charge
correlation function plotted in Fig. \ref{fig03}).

In the context of the present review focusing on dynamical screening
effects, the most interesting observation is however obtained from
an analysis of the spectral functions: The insulating state is formed,
as expected, from a splitting of the single half-filled surface band
into upper and lower Hubbard bands. The energetic separation of the
Hubbard bands is -- in the single orbital case -- a direct measure
of the effective local Coulomb interaction. However, this energy
is no longer set by the effective local Hubbard $U_0$ discussed
above. Indeed, non-local screening processes -- included within the
GW+DMFT scheme through the self-consistency over two-particle
quantities -- lead to a reduction of the Hubbard interaction to 
a smaller value $\mathcal{U}^{GW+DMFT}{(\omega=0)}$.
The frequency-dependent $\mathcal{U}^{GW+DMFT}(\omega)$
obtained within GW+DMFT for all four systems is plotted in Fig.\ref{fig03}.
The shape of this quantity is reminiscent of screened interactions, 
as calculated, e.g., within the cRPA\cite{PhysRevB.70.195104}, 
where retardation 
effects result from downfolding of high-energy degrees of freedom. 
The GW+DMFT $\mathcal{U}^{\rm GW+DMFT}(\omega)$ can be viewed as an effective 
interaction, where the dynamical character results from downfolding 
non-local degrees of freedom into a local quantity. 
At large frequencies, screening is not efficient and, hence, 
$\mathcal{U}^{\rm GW+DMFT}(\omega=\infty)=U_0$. On the other hand, the static 
value $\mathcal{U}^{\rm GW+DMFT}(\omega=0)$ can be significantly reduced (up 
to nearly a factor of 2 for Si(111):Pb). 
The transition between the
unscreened high frequency behavior and the static value takes 
place at an energy scale $\omega_0$ (plasmonic frequency) 
characteristic of the non-local charge fluctuations.
The last line in Table 1 summarises the static values of the
effective $\mathcal{U}^{GW+DMFT}(\omega=0)$.
These values are to be compared with the energetic separation of
upper and lower Hubbard band, that is, the insulating gap.
An estimate from the center of mass of the Hubbard bands yields values of 
$1.3$eV for Si(111):C, $0.8$eV Si(111):Si, 
$0.7$eV Si(111):Sn, and $0.5$eV for the
insulating solution of Si(111):Pb. 
These values are -- for the experimentally investigated systems,
and in particular Si:Sn which is the most studied one -- in excellent
agreement with experimental results.

Interestingly, the behavior of the dynamical interactions is strikingly
different between the different systems:
$\mathcal{U}(\omega)$ in Si(111):C [Si(111):Si] is [nearly] unaffected 
by non-local interactions and there is barely any screening. 
For Si(111):Sn and Si(111):Pb, however, 
the static values $\mathcal{U}^{\rm GW+DMFT}(\omega=0)$ are substantially
reduced compared to the 
onsite interaction: to $0.84$eV for Si(111):Sn and to $0.67$eV ($0.54$eV) 
for the insulating (metallic) solution for Si(111):Pb.
Plasmonic resonances at energies between $0.6$eV and $0.8$eV stress 
the importance of non-local interactions/charge-fluctuations for 
these systems.
It was proposed recently \cite{schueler} that the effect
of intersite interactions in extended Hubbard models for
surface systems could simply be described by a reduction of the
onsite interaction by the value of the nearest-neighbor one.
While full GW+DMFT calculations for an extended Hubbard model with
on- and intersite interactions only \cite{ayral} indeed confirm the
validity of this simple rule of thumb, the present more realistic
series demonstrates that this procedure would lead to a large
underestimation of the local interactions, when truly long-range
interactions are present in the original Hamiltonian.
In these cases, the efficiency of the non-local interaction in
screening the local ones seems to depend quite significantly
on the long-range interactions and/or the
underlying one-particle band structure.

\section{Perspectives}

From a technical point of view, fully dynamical GW+DMFT 
calculations remain a challenge, even nowadays, and 
full self-consistency could at present be achieved only
for the surface system thanks to the simplifications
that come along with a single orbital description.
Self-consistent GW+DMFT calculations for realistic
multi-orbital systems are thus an important goal for 
future work.
In this context, the orbital-separated scheme proposed
in the application to SrVO$_3$ reviewed above, where only a 
subset of low-energy states is treated within DMFT presents
appealing features.
For late transition metal oxides, and in particular
charge-transfer systems, interactions
between the correlated shell and the ligand electrons
are likely to be important for an accurate estimation
of the charge transfer energy and thus the whole electronic
structure, and exploring the performance of the GW+DMFT
scheme in this context is an interesting open problem.
One could still expect a perturbative treatment of the
intershell interactions to be sufficient, so that
explicitly including an intershell self-energy
$\Sigma_{pd}$ from GW in the orbital-separated scheme
discussed above seems a promising way.

At the same time, it would be most interesting if it
was possible to set up simpler 
schemes that could -- at least approximately --
reproduce the results of the full GW+DMFT calculations.
Two routes could be pursued:
\begin{itemize}
\item At the GW level,
is it possible to avoid full GW calculations, and to
base the combination with DMFT on a quasi-particle
self-consistent GW (qsGW) scheme, for example the one
by Kotani and Schilfgaarde \cite{schilfgaarde:226402} ?
\item At the DMFT level,
is it possible to avoid the solution of a local
problem with frequency-dependent interaction, replacing
it by a problem with static $U$ ?
\end{itemize}
Recent work provided important insights into the
first question \cite{jmt_qsGW, taranto}.
Ref. \cite{jmt_qsGW} argued that at least for certain
classes of materials -- the authors investigated iron
pnictide compounds -- nonlocal and dynamic 
contributions to the many-body self-energy are mostly
separable, so that a purely local self-energy correction
to a -- in some sense -- optimized one-particle band structure
would be sufficient for an accurate description.
The authors thus proposed a combination of quasi-particle
self-consistent GW \cite{schilfgaarde:226402} with DMFT.
Tests along these lines -- 
although without attempting to avoid double counting of
the local self-energies (see below) and based on a one-shot
GW calculation -- were performed in \cite{taranto}
for SrVO$_3$.

A subtle issue arises 
from the necessity of avoiding double 
counting of the local part of the GW self-energy,
when ``quasi-particlising'' the GW calculations.
Indeed, only the nonlocal part of the GW self-energy
enters the combined GW+DMFT self-energy,
which means that when a qsGW scheme is employed
the local part $\Sigma_{GW local}(\omega)= \sum_{k^{\prime}} GW$
has to be be explicitly substracted. 
This is not entirely trivial, since this
term acquires both k- and state-dependence through
the ``quasi-particlisation'' of its original
frequency-dependence. Indeed, the correction acting
on a quasi-particle state $\epsilon_{kn}$ reads
$\Sigma_{GW local}(\omega=\epsilon_{kn})$.
This term is responsible 
for the widening of the unoccupied bands discussed
above (see Sect. 6) and can thus by no means be
considered as negligible.
The implementation of efficient yet double-counting-free 
qsGW+DMFT schemes remains therefore an important
open challenge.

An optimistic answer can be given to
the second question, concerning
the need of using frequency-dependent interactions,
at least when only the very low-energy electronic
structure is of interest. 
Indeed, the same trick as discussed above for
reducing dynamical Hubbard models to static ones
at the price of a renormalisation of the one-body
part of the Hamiltonian can be applied also to a
GW calculation corrected for its local term
(``non-local GW''), ``quasi-particlised'' or not.
On the other hand, the gain in computational
cost obtained by this procedure is not tremendous
in this case, since at least within the BFA the 
solution of a dynamical local model can anyhow be 
achieved at the cost of the solution of the 
corresponding static model.

Finally, the GW+DMFT construction also introduces
some non-locality into the otherwise purely
local self-energy (and polarisation) of (extended)
DMFT. The first realistic GW+DMFT calculations 
for SrVO$_3$ have evidenced a crucial effect of this
non-local self-energy on the {\it unoccupied} part
of the Kohn-Sham band structure. A widening of the
band due to the exchange interaction, which -- for
the empty states -- is not well described by the
Kohn-Sham potential leads to a substantial reinterpretation
of the electronic structure. Nevertheless, this
correction is essentially static and given by the
Fock exchange, the truly frequency-dependent
non-local part of the self-energy correction is
small.
A recent systematic study on the two-dimensional
extended Hubbard model \cite{ayral} 
has in fact shown that unless considering a
system close to the charge-ordering transition,
these non-local self-energy effects are tiny. To capture
the strong k-dependence of the self-energies in the 
immediate vicinity of the Mott transition, experimentally
observed e.g. in the form of highly k-dependent
variations of the effective masses in doped Mott
insulators, extending
GW+DMFT to include not only fluctuations in the
charge but also in the spin channel seems necessary.
This is yet another promising topic for future
work.

\section{Conclusions}

We have reviewed a series of recent papers dealing
with dynamical screening of the Coulomb interactions
in materials with correlated electrons.
Quite generally, dynamically screened interactions
arise as a ``representative'' within a restricted subspace 
of the full long-range Coulomb interactions in the 
original Hilbert space. In practice, it appears to be useful
to disentangle two different mechanisms: 
(1) screening by high-energy degrees of
freedom, when the original Coulomb Hamiltonian is downfolded
to a low-energy effective Hamiltonian, (2) screening by
non-local degrees of freedom, when a Hamiltonian with long-range
interaction is backfolded into one with purely local interactions.
The cRPA provides a simple and transparent description 
of the first mechanism,
and the energetic separation of the different degrees of freedom
can probably in many cases justify the neglect of vertex corrections.
This is less obvious in the case of spatial downfolding, where
both ``backfolded'' and retained degrees of freedom live on the
same energy scales. The GW+DMFT scheme provides an elegant
construction of effective local interactions, while {\it including
a local vertex in a non-perturbative manner}.

This review has discussed the effects of dynamical interactions,
specifically on the examples of the ternary transition metal
compound SrVO$_3$ and the iron pnictide BaFe$_2$As$_2$. It was
argued that the dynamical nature of the Hubbard interaction
leads to many-body satellites in the spectral function
(corresponding to plasmons, particle-hole excitations or more
general bosonic excitations), and that the corresponding 
transfer of spectral weight leads to additional renormalisations
in the low-energy electronic structure.

For SrVO$_3$, which was the first compound treated within
GW+DMFT in 
a fully dynamical manner (albeit selfconsistent only at the DMFT level),
the non-local self-energy effects introduced by this
scheme moreover lead to profound modifications of the
unoccupied spectra, as compared to simple LDA+DMFT
calculations. In particular, the peak at 2.7 eV seen in
inverse photoemission experiments was identified as an
$e_g$ feature, while the energy scale of the upper Hubbard
($\sim$ 2 eV) makes this feature hard to separate from the
quite dispersive unoccupied t$_{2g}$ band states.
These findings require a reinterpretation of early LDA+DMFT
calculations using a t$_{2g}$-model for this compound, 
where the peak at 2.7 eV was commonly interpreted as
an upper Hubbard band of $t_{2g}$ character.
They do reconcile however DMFT-based electronic structure
calculations with cluster model calculations for SrVO$_3$
\cite{mossanek:033104}.
The results are furthermore encouraging, since this first
benchmark on SrVO$_3$ confirms the ability
of the GW+DMFT approach to describe simultaneously Hubbard 
bands, higher energy satellite structures and corrected energy 
gaps, in an {\it ab initio} fashion.
Quite generally, materials with a
\textquotedblleft double LDA failure\textquotedblright , 
an inappropriate description of 
correlated states, and deficiencies of LDA for the
more itinerant states, such as an underestimated
``pd-gap'', can be treated within this scheme. 

Finally, we have reviewed an illustrative example 
of spatial downfolding within the self-consistent
GW+DMFT scheme for the low-energy electronic structure
of two-dimensional systems of atoms adsorbed on the Si(111) 
surface. Here, the frequency-dependence of the interactions 
stemming from downfolding higher energy degrees of freedom
is weak (and thus neglected), but non-local screening
leads to (system-dependent) reductions of the effective 
local interaction of up to nearly 50$\%$.

These examples demonstrate the virtues of the GW+DMFT scheme:
\begin{itemize}
\item GW+DMFT is entirely formulated in the Green's function
language, even at the one-body level. In this way, the
theory is conceptually
double counting-free, since it avoids the mismatch
between the Green's function-based description of DMFT and the
density-based one of DFT that is inherent to any combined
DFT+DMFT scheme.
\item It deals directly with the long-range
Coulomb interactions, and the effective local Hubbard
interactions arise only as intermediate auxiliary
quantities. This has several consequences: first, it allows
for a truly {\it ab initio} description of the Coulomb interactions,
including dynamical screening effects (and the associated
transfers of spectral weight, plasmon replicae etc) as described above.
Second, when intersite interactions become important, as e.g. in the
surface systems described above, new instabilities (here towards 
charge-ordered phases) can appear.
\item It retains the non-perturbative
character of dynamical mean field theory, thus avoiding limitations
due to a truncation of the perturbation series. This latter point
is essential to ensure the scheme to be equally appropriate in
the weak, strong and intermediate coupling regimes.
\item Finally, the framework of the orbitally separated GW+DMFT
scheme described above,
where perturbative corrections to the LDA band structure are
applied to ligand or conduction band states allows for an
accurate description of the electronic structure on larger energy
scales than in conventional methods. Indeed, within conventional
LDA+DMFT techniques, there is no correction to the LDA for
ligands or conduction band states (other than shifts resulting
from suppressed hybridisations with the correlated shell).
As demonstrated on the example of SrVO$_3$, such effects can
be crucial even for a qualitative assessment of electronic
states on energy scales of 2 or 3 eV.
\end{itemize}
We close this review by stressing again that the most important
lesson to be learnt from the recent calculations is probably the 
finding that without addressing dynamical screening effects 
a ``first principles'' description
is not even possible for the one-body part of the Hamiltonian:
even in cases where a good separation of energy scales ($\omega_P$ $>>$
any other energy scale in the system) allows for an accurate 
mapping onto a static Hubbard model with effective local interaction,
the bosonic renormalisations are crucial for assessing the one-particle
Hamiltonian in a quantitative way. On the other hand, it
is well-known -- and has been
demonstrated again and again by numerous examples over the last
years -- that even tiny differences in the one-particle part of
the Hamiltonian can lead, in a strongly correlated situation, to
very different physical behaviors. Indeed, quantitative differences at
the one-particle level can easily make the difference between
a metal or an insulator \cite{pavarini:176403, martins}.
Even on the level of simple one-orbital models, inclusion of
dynamical screening changes the critical interaction for the
Mott transition, thus shifting the metal-insulator transition line
\cite{casula_effmodel}.

When are dynamical screening effects stemming from higher energy
degrees of freedom large? The strength of the
renormalisations induced by dynamical screening depends on the
ratio of the electron-boson coupling strength over the plasma
frequency, so that strong effects can be expected if the
plasma frequency is small and the coupling large.
Within the class of t$_{2g}$ transition metal oxides, the
plasma frequencies vary substantially less than the couplings.
Indeed, the coupling strength is given by the difference of
bare and (partially) screened interaction, 
$\Delta \mathcal{U} = \mathcal{U}(\infty) - \mathcal{U}(0)$.
Typical values vary from $\Delta \mathcal{U}\sim $ 12 eV
for SrVO$_3$ to $\sim $ 8 eV for 5d oxides, such as Sr$_2$IrO$_4$
\cite{martins}.
The importance of the effects can thus be expected to 
decrease from 3d to 4d and 5d oxides.
On the other hand, 4d and 5d orbitals are more extended than
3d orbitals, and intersite interactions are therefore larger%
\footnote{The ratio of intersite to onsite interactions for
5d oxides can be as large as 50 $\%$ \cite{pseth}}. 
Backfolding of long-range interactions as in
the case of the surface systems discussed above is therefore
expected to generate larger corrections than in the 3d oxides.

For late transition metal oxides, where small charge-transfer
energies and entangled d- and ligand states lead to additional
complications, these questions are still largely unexplored.
The role of intershell (correlated to ligand shell) interactions,
of long-range exchange, or of corrections to the LDA estimate 
of the d-p hybridisations are ingredients that are still
awaiting systematic investigations.
Progress is rapid, but it is also clear that there is still
some way to go until first principles many-body techniques
will have realised their full predictive potential, suitable
to serve materials scientists in designing new unknown
materials.

\section{Acknowledgments}

This review summarises elements of the series of works
\cite{PhysRevB.70.195104, PhysRevB.86.165105, 
casula_effmodel, 
PhysRevB.85.035115, werner_Ba122, 
gwdmft, ayral, ayral-prb, 
jmt_svo, jmt_svo2, hansmann} 
that I carried out together with 
different coauthors. I thank
F. Aryasetiawan, T. Ayral, M. Casula, A. Georges, P. Hansmann, 
H. Jiang, A. Millis, T. Miyake, A. Rubtsov,
J.M. Tomczak, L. Vaugier, P. Werner
most warmly for the fruitful and enjoyable collaborations, 
all the way from the first proposal of the GW+DMFT scheme \cite{gwdmft}
until its first dynamical implementations for realistic materials
\cite{jmt_svo, hansmann}.
I furthermore acknowledge useful discussions with M. Ferrero, M. Imada, 
M.I. Katsnelson, A.I. Lichtenstein, O. Parcollet, L. Pourovskii,
A. van Roekeghem, S. Sakai, and
G. Sawatzky on various aspects related to the topics of this review.
\\

This work was supported by the French ANR under projects SURMOTT 
and PNICTIDES, and IDRIS/GENCI under project 139313.

\section*{Appendix A: Effective low-energy Hamiltonian
incorporating renormalisations due to dynamical
screening -- the electronic polaron}

It has been discussed above how dynamical Hubbard interactions
can be incorporated into the many-body description by LDA+DMFT,
and that they are an integral part of the combined GW+DMFT
scheme. As seen on the above examples, even
if the characteristic screening frequency (plasma frequency)
is much larger than other relevant low-energy energy scales
of the system (bandwidth and static Hubbard $U$),
dynamical screening leads to 
substantial renormalisations of the low-energy electronic 
structure. One can thus ask the question if in this antiadiabatic 
limit 
it is possible to construct a low-energy effective Hamiltonian
{\it with static Hubbard interactions} that reproduces the
spectral properties of the dynamical problem in a low-energy
window around the Fermi energy.

The answer is yes, and has been worked out explicitly in \cite{casula_effmodel},
where it was shown that in the antiadiabatic limit
the dynamical effect of the interaction can be captured
by a simple rescaling procedure of the {\it one-particle
part} of the Hamiltonian.
In the case of an Hamiltonian containing only low-energy
degrees of freedom treated as ``correlated'', the
effect amounts to a simple scaling factor on the one-particle
part of the Hamiltonian which is multiplied by (\ref{Z_B}).
The derivation of this result is given in \cite{casula_effmodel},
and relies on a Lang-Firsov transformation to a polaronic
Hamiltonian, where the coupling of electrons and plasmons
(or bosonic particle-hole excitations) is eliminated at the
price of passing to more complex (polaronic) degrees of
freedom and then projecting onto the low-energy space 
containing no plasmon- or particle-hole contributions.

The final result is a downfolded many-body Hamiltonian
that has the form of an extended Hubbard model, where the
one-particle hoppings have been renormalised (for simplicity
we only give here the single-orbital case, but the
generalisation to multi-orbital systems, also in the
presence of further uncorrelated orbitals was given in 
\cite{casula_effmodel}):
\begin{eqnarray}
\hat{H}= Z_B \sum_{ij\sigma}t_{ij}\left(\hat{c}^\dagger_{i\sigma} \hat{c}_{j\sigma}+{\rm h.c.}\right)
+
H_{int}[\mathcal{U}(\omega=0)]
\label{exthubbard}
\end{eqnarray}
Here, $\hat{c}^\dagger_{i\sigma}$($ \hat{c}_{i\sigma}$) creates (annihilates) 
an electron with spin $\sigma$ 
at lattice site $i$; 
$t_{ij}$ is the hopping amplitude between the Wannier orbitals on
lattice sites $i$ and $j$, and $H_{int}[U]$
$\mathcal{U}(\omega=0)]$ 
is the interaction term of (extended) Hubbard form (in the
multi-orbital case possibly including Hund's rule coupling etc.).
The resulting prescription is thus rather simple:
\begin{itemize}
\item
the one-particle part stems from a one-particle-downfolding
procedure, supplemented by a rescaling by $Z_B$.
\item
the interaction is the zero-frequency limit of the
dynamical Hubbard $U$ (as e.g. calculated within the cRPA).
In the one-orbital case, this is simply the matrix element
\begin{eqnarray}
U  = \langle \phi_{R} \phi_{R} | W^{\rm rest}(0) | \phi_{R} \phi_{R}\rangle \label{umatrix}.
\end{eqnarray}
\end{itemize}
The problem with dynamical interactions can thus be mapped
back again onto a static problem, but with renormalised
fermions, corresponding to the mass enhancement due to the
coupling to the bosonic degrees of freedom. In analogy
to the electron-phonon coupling problem%
\footnote{For the electron-phonon problem, there exists
an extended literature of exploring under which conditions
the problem can be mapped onto an effective (possibly
negative-U) Hubbard model, see e.g.~\cite{PhysRevB.52.4806,
Capone2010}.
}, the resulting
fermionic charge carriers of enhanced mass -- electrons
dressed by screening bosons (plasmons or particle-hole
excitations) -- are called ``electronic polarons''
\cite{Hedin-1965, Macridin, casula_effmodel}.

\section*{Appendix B: Derivation of Double Counting
Term in the Presence of Electron-Boson Coupling
}

In this appendix, we explore the consequences of the presence
of the bosonic degrees of freedom for the double counting terms.
In particular, we show that the standard arguments used for
deriving the double counting terms within LDA+U or LDA+DMFT
lead, in this case, to the same conclusions. In particular,
despite of the {\it bare} interactions entering the two-body
term, the {\it screened} interactions should be used within
the double counting correction term.

For simplicitly, we consider the case of a one-orbital model.
The Hamiltonian of a Hubbard-model coupled to a plasmon
mode reads:
\begin{eqnarray}
H = H_0 + V \sum_i n_{i\uparrow} n_{i\downarrow} 
+ 
\sum_i \omega_0 b_i^{\dagger} b_i + \lambda (b_i^{\dagger} + b_i) 
(n_{i\uparrow} + n_{i\downarrow} -1)
\end{eqnarray}
The quadratic (in the fermionic operators) part of the
Hamiltonian can be obtained from DFT-LDA, a procedure
which in the general case of coexisting ``correlated''
and ``uncorrelated'' orbitals needs to be corrected
by a suitable double-counting term:
\begin{eqnarray}
H_0 = H_{LDA} - V_{dc} (n_{i\uparrow} + n_{i\downarrow})
\end{eqnarray}
In the absence of the plasmon mode ($\omega_0=0$,
$\lambda=0$), the usual ``LDA+U'' or ``LDA+DMFT''
arguments would correspond to 
\begin{eqnarray}
V_{dc} = V (N - \frac{1}{2})
\end{eqnarray}
with the {\bf bare} Coulomb interaction V.

However, adding the plasmon requires an additional
double counting corresponding to the {\it DFT-LDA}
treatment of the plasmon.
Let us forget for the moment that we use the
LDA as an approximation to the exact density
functional. The following argument applies
thus to a Hamiltonian $H_0$ constructed from the
{\it exact} density functional.
Then we argue that the DFT ground
state energy is {\bf exact}, any corrections
should only change {\it spectral properties}.
This leads to an additional double counting
term such that the average value
\begin{eqnarray}
\langle
\sum_i \omega_0 b_i^{\dagger} b_i + \lambda (b_i^{\dagger} + b_i) 
(n_{i\uparrow} + n_{i\downarrow} -1)
- \Delta V_{dc} (n_{i\uparrow} + n_{i\downarrow})
\rangle
\end{eqnarray}
vanishes at mean field level.
This can be achieved by
choosing
\begin{eqnarray}
\Delta V_{dc} =
\lambda \langle
(b_i^{\dagger} + b_i) 
\rangle
\end{eqnarray}
The mean value
\begin{eqnarray}
\langle
\lambda (b_i^{\dagger} + b_i) 
\rangle
\end{eqnarray}
can be evaluated and gives
\begin{eqnarray}
\langle
\lambda (b_i^{\dagger} + b_i) 
\rangle
= - \frac{2 \lambda^2}{\omega_0}
\langle (n_{i\uparrow} + n_{i\downarrow} -1) \rangle
\end{eqnarray}
Thus we obtain for the additional
double counting:
\begin{eqnarray}
\Delta V_{dc} =
- \frac{2 \lambda^2}{\omega_0}
\langle (n_{i\uparrow} + n_{i\downarrow} -1) \rangle
\end{eqnarray}
Interestingly, for the one-band model,
where the Holstein coupling $\lambda$
is simply related to the dynamical
interaction by
\begin{eqnarray}
\lambda^2 = \frac{(V-U_0) \omega_0}{2}
\end{eqnarray}
and thus
\begin{eqnarray}
V - \frac{2 \lambda^2}{\omega_0} = U_0
\end{eqnarray}
The plasmon double counting thus
eliminates the bare interaction from the
total double counting.
Analogous arguments can be found in the
multi-orbital case, if the screening
bosons couple only to the total charge.

We thus arrive at the attractive result
that the double counting terms for
LDA+$\mathcal{U}(\omega)$+DMFT should be
identical to the ones used for standard
LDA+U or LDA+DMFT.

\section*{Appendix C: Derivation of GW+DMFT from
a Free Energy Functional}

In this Appendix we review the derivation of the GW+DMFT
scheme from a functional point of view.
The discussion follows closely the original derivation
in \cite{gwdmft}.

As noted in \cite{Almbladh_1999, chitra},
the free energy of a solid can be viewed as a functional 
$\Gamma[G,W]$ of the
Green's function $G$ and the screened Coulomb interaction $W$.
The functional $\Gamma$ can trivially be split into a Hartree
part $\Gamma_H$ and a many body correction $\Psi$, which
contains all corrections beyond the Hartree approximation~:
$\Gamma =  \Gamma_H + \Psi$. The Hartree part can be given
in the form
\begin{eqnarray}
\Gamma_H[G,W]
&=&Tr\ln G-Tr[({G_H}^{-1}-G^{-1})G]
\nonumber
\\
&-&\frac{1}{2}Tr\ln W+\frac{1}%
{2}Tr[({V_q}^{-1}-W^{-1})W]
\label{LW}%
\end{eqnarray}
The $\Psi$-functional
is the sum of all skeleton
diagrams that are irreducible with respect to both, one-electron
propagator and interaction lines.
 $\Psi[G,W]$ has the following properties:
\begin{eqnarray}\label{eq:def_sigma_P} 
\frac{\delta \Psi}{\delta G} = \Sigma^{xc}
\nonumber
\\
\frac{\delta \Psi}{\delta W} = P.
\end{eqnarray}
The $\Psi$ functional was first derived in \cite{Almbladh_1999}.
A detailed discussion in the context of extended DMFT can
be found in \cite{chitra}.

The GW approximation consists in retaining the first order
term in the screened interaction $W$ only, thus 
approximating the $\Psi$-functional by
\begin{eqnarray}
\Psi[G,W] = -\frac{1}{2} Tr(GWG).
\end{eqnarray}
We then find trivially
\begin{eqnarray}
\Sigma = \frac{\delta \Psi}{\delta G} = -G W
\end{eqnarray}
\begin{eqnarray}
P = \frac{\delta \Psi}{\delta W} = G G.
\end{eqnarray}
 
Extended DMFT, on the other hand, would calculate all
quantities derived from this function from a local
impurity model, that is, one can formally write
\begin{eqnarray}\label{Psi-edmft}
\Psi\,=\,\Psi_{imp}[\GRR,\WRR].
\end{eqnarray}

In \cite{gwdmft}, an approximation to the $\Psi$ functional
was constructed that corresponds to the combined GW+DMFT
scheme. It approximates 
the $\Psi$ functional as a direct combination of local
and non-local parts from GW and extended DMFT respectively:

\begin{eqnarray}\label{Psi}
\Psi\,=\,\Psi_{GW}^{\rm{non-loc}}[\GRRp,\WRRp]
+\Psi_{imp}[\GRR,\WRR]
\end{eqnarray}
 
More explicitly, the non-local part of the GW+DMFT
$\Psi$-functional is given by
\begin{eqnarray}
\Psi_{GW}^{\rm{non-loc}}[\GRRp,\WRRp]
= \Psi_{GW}[\GRRp,\WRRp] - \Psi_{GW}^{\rm{loc}}[\GRRp,\WRRp]
\end{eqnarray}
while the local part is taken to be an impurity model 
$\Psi$ functional.
Following (extended) DMFT, this onsite part of the functional is generated
from a local {\it quantum impurity problem} (defined on a single
atomic site).
The expression for its free energy functional
$\Gamma_{imp}[G_{imp},W_{imp}]$
is analogous to (\ref{LW}) with 
${\cal G}$ replacing $\GH$ and ${\cal U}$ replacing $V$~:
\begin{eqnarray}
\Gamma_{imp}[G_{imp},W_{imp}]
&=&Tr\ln G_{imp}-Tr[({\cal G}^{-1}-G_{imp}^{-1})G_{imp}]
\nonumber
\\
&-&\frac{1}{2}Tr\ln W_{imp}+\frac{1}%
{2}Tr[({\cal U}^{-1}-W_{imp}^{-1})W_{imp}]
\nonumber
\\
&+&\Psi_{imp}\lbrack G_{imp},W_{imp} \rbrack \label{LWimp}%
\end{eqnarray}
The impurity quantities $G_{imp},W_{imp}$ can thus be
calculated from the
effective action:
\begin{eqnarray}\label{eq:action}\label{Simp}
&S&=\int d\tau 
d\tau' \left[ -
\sum 
c^{\dagger}_{L}(\tau) {\cal G}^{-1}_{LL'}(\tau-\tau')c_{L'}(\tau')
\right.
\\ \nonumber
&+&\frac{1}{2}
\left.
\sum 
:c^\dagger_{L_1}(\tau)c_{L_2}(\tau):\cU_{L_1L_2L_3L_4}(\tau-\tau')
:c^\dagger_{L_3}(\tau')c_{L_4}(\tau'): \right]
\hskip-1cm
\end{eqnarray}
where the sums run over all orbital indices $L$.
In this expression, $c_L^{\dagger}$ is a creation operator associated with orbital $L$
on a given sphere, and the double dots denote normal ordering (taking care of
Hartree terms).

The construction (\ref{Psi})
of the $\Psi$-functional is the only ad hoc
assumption in the GW+DMFT approach. The explicit form of the
GW+DMFT equations follows then directly from the functional relations
between the free energy, the Green's function, the screened
Coulomb interaction etc.
 Taking derivatives of (\ref{Psi}) as in (\ref{eq:def_sigma_P}) it is seen
that the complete self-energy and polarization operators read:
\begin{eqnarray}\label{Sig_a}
\Sigma^{xc}(\vk,\iomn)_{LL'} &=& \Sigma_{GW}^{xc}(\vk,\iomn)_{LL'}
\\
&-& \sum_\vk \Sigma_{GW}^{xc}(\vk,\iomn)_{LL'}
+ [\Sigma^{xc}_{imp}(\iomn)]_{LL'}
\nonumber
\\
\label{P_a}
P(\vq,\inun)_{\a\b} &=& P^{GW}(\vq,\inun)_{\a\b}
\\
&-& \sum_\vq P^{GW}(\vq,\inun)_{\a\b}+ P^{imp}(\inun)_{\a\b}
\nonumber
\end{eqnarray}

The meaning of (\ref{Sig_a}) is transparent:
the off-site part of the self-energy is taken from
the GW approximation, whereas the onsite part is calculated
to all orders
from the dynamical impurity model.
This treatment thus goes beyond usual E-DMFT, where
the lattice self-energy and polarization are just taken
to be their impurity counterparts.
The second term in
(\ref{Sig_a}) substracts the onsite component of the GW
self-energy thus avoiding double counting.
At self-consistency this term can be rewritten as:
\begin{equation}\label{Sig_correction}
\sum_\vk \Sigma_{GW}^{xc}(\tau)_{LL'}= - \sum_{L_1L_1'}
W^{imp}_{LL_1L'L'_1}(\tau) G_{L'_1L_1}(\tau)
\end{equation}
so that
it precisely substracts the contribution of the GW diagram to
the impurity self-energy. Similar considerations apply
to the polarization operator.

We now outline the iterative loop which determines $\cG$ and $\cU$ self-consistently
(and, eventually, the full self-energy and polarization operator):
\begin{itemize}
\item The impurity problem (\ref{Simp}) is solved, for a given choice of $\cG_{LL'}$ and
$\cU_{\a\b}$: the ``impurity'' Green's function
\begin{equation}
G_{imp}^{LL'}\equiv - \langle T_\tau c_L(\tau)c^+_{L'}(\tau')\rangle_S
\end{equation}
is calculated, together with the impurity self-energy
\begin{equation}
\Sigma^{xc}_{imp}\equiv\delta\Psi_{imp}/\delta G_{imp}=\cG^{-1}-G_{imp}^{-1}.
\end{equation}
The two-particle correlation function
\begin{equation}
\chi_{L_1L_2L_3L_4}=\langle :c^\dagger_{L_1}(\tau)c_{L_2}(\tau):
:c^\dagger_{L_3}(\tau')c_{L_4}(\tau'):\rangle_S 
\end{equation}
must also be evaluated.
\item
The impurity effective interaction is constructed as follows:
\begin{equation}\label{eff_inter}
W_{imp}^{\a\b} = \cU_{\a\b} -
\sum_{L_1\cdots L_4}\sum_{\gamma\delta} \cU_{\a\gamma}
O^{\gamma}_{L_1L_2}
\chi_{L_1L_2L_3L_4} [O^{\delta}_{L_3L_4}]^* \cU_{\delta\b}
\end{equation}
where 
$O_{L_1L_2}^\a\equiv\langle\phi_{L_1}\phi_{L_2}|B^\a\rangle$
is the overlap matrix between two-particle states and products
of one-particle basis functions.
The polarization operator of the impurity problem is then obtained as:
\begin{equation} 
P_{imp}\equiv -2\delta\Psi_{imp}/\delta W_{imp} = \cU^{-1}-W_{imp}^{-1},
\end{equation} 
where all  
matrix inversions are performed in the two-particle basis
$B^\a$
(see the discussion in \cite{gwdmft_proc1, gwdmft_proc2}).
\item
From Eqs.~(\ref{Sig_a}) and (\ref{P_a})
the full $\vk$-dependent Green's function $G(\vk,\iomn)$ and
effective interaction $W(\vq,\inun)$ can be constructed. The self-consistency condition
is obtained, as in the usual DMFT context, by requiring that the onsite
components of these quantities coincide with $G_{imp}$ and $W_{imp}$. In practice, this
is done by computing the onsite quantities
\begin{eqnarray}\label{Glocal}
G_{loc}(\iomn) &=& \sum_\vk [\GH^{-1}(\vk,\iomn) - \Sigma^{xc}(\vk,\iomn) ]^{-1}\\
\label{Wlocal}
W_{loc}(\inun) &=& \sum_\vq [V_{\vq}^{-1} - P(\vq,\inun)]^{-1}
\end{eqnarray}
and using them to update the
Weiss dynamical mean field ${\cal G}$
and the impurity model interaction ${\cal U}$ according to:
\begin{eqnarray}\label{update}
\cG^{-1} = G_{loc}^{-1} + \Sigma_{imp}
\\
\label{updateU}
\cU^{-1} = W_{loc}^{-1} + P_{imp}
\end{eqnarray}
\end{itemize}
The set of equations (28) to (36) is iterated until self-consistency.


\begin{thebibliography}{10}
\bibitem{silicon} The structure with silicon as an adatom does not exist. It is calculated/presented as a hypothetical structure to complete the series - the pure silicon 111 surface is not stable in the $\sqrt{3}\times\sqrt{3}$ phase, but reorders to a very large $7\times 7$ unit cell.
\end{thebibliography}
\end{document}